\documentclass[aps,10pt,prl,twocolumn,letterpaper,superscriptaddress]{revtex4-1}

\usepackage{graphicx} % Include figure files
\usepackage{bm} % bold math
\usepackage{amsmath}
\usepackage{xcolor}

\definecolor{PRLblue}{rgb}{0.18,0.18,0.57}
\usepackage[colorlinks=true,allcolors=PRLblue]{hyperref} % add hypertext capabilities
\graphicspath{ {./}{figs/}}
\usepackage{fancyhdr}
\begin{document}
\title{Effect of Localization on Photoluminescence  and Zero-Field Splitting of Silicon Color Centers}

\author{Vsevolod Ivanov}
\affiliation{Molecular Foundry, Lawrence Berkeley National Laboratory, Berkeley, CA 94720, USA}
\affiliation{Accelerator Technology and Applied Physics Division, Lawrence Berkeley National Laboratory, Berkeley, CA 94720, USA}
\author{Jacopo Simoni}
\affiliation{Molecular Foundry, Lawrence Berkeley National Laboratory, Berkeley, CA 94720, USA}
\author{Yeonghun Lee}
\affiliation{Molecular Foundry, Lawrence Berkeley National Laboratory, Berkeley, CA 94720, USA}
\affiliation{Department of Electronics Engineering, Incheon National University, Incheon 22012, Republic of Korea}
\author{Wei Liu}
\affiliation{Accelerator Technology and Applied Physics Division, Lawrence Berkeley National Laboratory, Berkeley, CA 94720, USA}
\author{Kaushalya Jhuria}
\affiliation{Accelerator Technology and Applied Physics Division, Lawrence Berkeley National Laboratory, Berkeley, CA 94720, USA}
\author{Walid Redjem}
\affiliation{Department of Electrical Engineering and Computer Sciences, University of California, Berkeley, CA 94720, USA}
\author{Yertay Zhiyenbayev}
\affiliation{Department of Electrical Engineering and Computer Sciences, University of California, Berkeley, CA 94720, USA}
\author{Christos Papapanos}
\affiliation{Department of Electrical Engineering and Computer Sciences, University of California, Berkeley, CA 94720, USA}
\author{Wayesh Qarony}
\affiliation{Department of Electrical Engineering and Computer Sciences, University of California, Berkeley, CA 94720, USA}
\author{Boubacar Kante}
\affiliation{Department of Electrical Engineering and Computer Sciences, University of California, Berkeley, CA 94720, USA}
\affiliation{Accelerator Technology and Applied Physics Division, Lawrence Berkeley National Laboratory, Berkeley, CA 94720, USA}
\author{Arun Persaud}  % ORCID: 0000-0003-3186-8358
\affiliation{Accelerator Technology and Applied Physics Division, Lawrence Berkeley National Laboratory, Berkeley, CA 94720, USA}
\author{Thomas Schenkel}
\affiliation{Accelerator Technology and Applied Physics Division, Lawrence Berkeley National Laboratory, Berkeley, CA 94720, USA}
\author{Liang Z. Tan}
\affiliation{Molecular Foundry, Lawrence Berkeley National Laboratory, Berkeley, CA 94720, USA}

\begin{abstract}	
	
The study of defect centers in silicon has been recently reinvigorated by their potential applications in optical quantum information processing. A number of silicon defect centers emit single photons in the telecommunication $O$-band, making them promising building blocks for quantum networks between computing nodes. The two-carbon G-center, self-interstitial W-center, and spin-$1/2$ T-center are the most intensively studied silicon defect centers, yet despite this, there is no consensus on the precise configurations of defect atoms in these centers, and their electronic structures remain ambiguous. Here we employ \textit{ab initio} density functional theory to characterize these defect centers, providing insight into the relaxed structures, bandstructures, and photoluminescence spectra, which are compared to experimental results. Motivation is provided for how these properties are intimately related to the localization of electronic states in the defect centers. In particular, we present the calculation of the zero-field splitting for the excited triplet state of the G-center defect as the structure is linearly interpolated from the A-configuration to the B-configuration, showing a sudden increase in the magnitude of the $D_{zz}$ component of the zero-field splitting tensor. By performing projections onto the local orbital states of the defect, we analyze this transition in terms of the symmetry and bonding character of the G-center defect which sheds light on its potential application as a spin-photon interface. 

\end{abstract}

\maketitle

\section{I. Introduction}

%Introduction/Motivation

Point defect centers are receiving increasing attention due to their potential applications for quantum information science (QIS). Point defects in silicon hold a number of advantages in this regard, exhibiting photon emission in the telecommunication band \cite{tcenter-spin-photon, gali-GC}, long electron spin coherence times\cite{Tyryshkin_2006}, and narrow linewidths \cite{defect-engineering}, with promise to enable large-scale integration of quantum communication between local quantum computer nodes\cite{Awschalom2018}. Fabrication of silicon based electronics is widespread, and there are various synthesis processes and quality control measures that have been developed for atomic level control of silicon-based devices\cite{QIS-review, intro-Durand, intro-Zhang}. Furthermore, most modern telecommunications and computing devices are based on silicon, allowing any prospective silicon-based quantum devices to be more easily integrated with existing technologies. Potential applications of such color centers for quantum networks can leverage the established silicon-based manufacturing processes of integrated electronics and photonics platforms.

Work on understanding the PL properties of silicon has been ongoing for nearly half a century. In that time, a general understanding has emerged of the various point defects that can arise in silicon, the techniques for generating them through radiation damage and annealing, as well as their structures and vibronic properties. More recently, the focus has turned to isolating high-quality defect centers with narrow PL linewidths, with ongoing efforts to create silicon-based devices that can generate indistinguishable photons, couple spin and photon degrees of freedom, and various other properties needed for QIS applications \cite{intro-Durand, intro-Zhang, QIS-review}.

Despite the promise of silicon defect centers for QIS applications, there exists no firm consensus in regard to their structural and electronic properties. In particular, a deeper understanding is needed of the electronic defect levels within the bulk silicon gap and the nature of the local states corresponding to these levels. In this work, we present a theoretical study of the electronic properties of the G-center \cite{gali-GC, GC1,GC2,GC3,Gcenter-HSE}, W-center \cite{WC-theory-Vdiscov, WC-theory-V1, WC-theory-V2}, and T-center \cite{sinead-TC, tcenter-safonov}, which have been experimentally observed to have a dominant zero-phonon line within or near the telecommunication bands for low loss transmission of photons through optical fibers. Using a combination of first-principles calculations and tight-binding models, we present calculations of the band structures, photoluminescence spectra, and zero-field splitting (ZFS) parameters, and discuss them in relation to recent experimental results. We discuss these properties in the context of the local defect structure and show how they can be affected dramatically by the localization of the defect electronic states.

This work is structured as follows: in Section II, we summarize the progress in understanding the structures of these various defect centers and then compare with our first-principles calculations in Section III. Section IV outlines the procedure for computing the photoluminescence and presents the spectra for the three defect structures, which are compared with experiments. In Section V, we focus on the ZFS in the excited triplet state of the G-center and discuss aspects of the electronic structure and localization that enhance the ZFS. Finally in Section VI, we conclude with a perspective on the field and how these silicon point defects may aid in the development of QIS technologies.

\section{II. Background}

Understanding the details of the electronic structure of silicon point defects is integral to the development of new QIS technologies based on silicon. In particular, there is a need for defects that have narrow linewidths within or near the telecommunication bands, and can couple photons to spin degrees of freedom, making the zero phonon line (ZPL), ZFS, and photoluminescence (PL) crucial properties for computation. These electronic properties are quite sensitive to the structure of the defects \cite{WC-theory-Vdiscov,Gcenter-HSE, gali-GC, GC1}, which can have several possible configurations with similar energetic and electronic structure.
%Despite this, the exact atomic positions are often omitted from publications. 
In order to be precise about which defects are being computed, the relaxed atomic positions are provided in the supplemental material \cite{supplement}.

\begin{figure}[bp]
	%\captionsetup{justification=raggedright}
	\includegraphics[width=1.0\columnwidth]{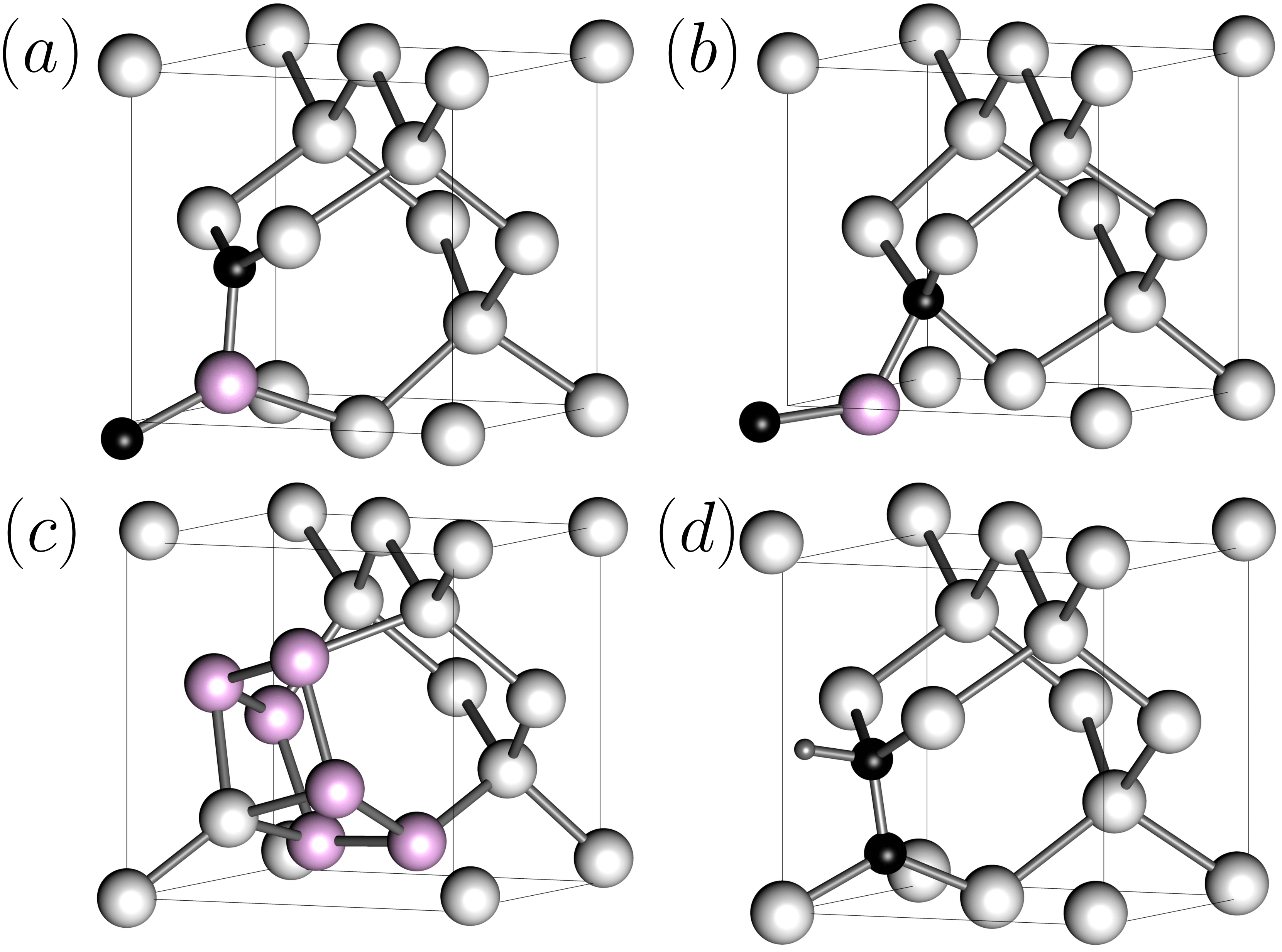}
	\caption[Defect structures in silicon]{Defect structure within the silicon unit cell for (a) G-center type A, (b) G-center type B, (c) W-center type V, and (d) T-center. Silicon atoms are colored white, carbon atoms black, hydrogen atoms dark grey, and silicon atoms part of the defect cluster in pink.}
	\label{struct}
\end{figure}

One such defect is the the silicon W-center, which was originally identified as a narrow band at 1018 meV in the PL spectrum of silicon that had been exposed to neutron or ion radiation damage\cite{1989_GDavies_PR}. Under uniaxial compression, the perturbation of the emission line is highly non-linear, and was used to identify the transition as occurring between non-degenerate spinless states in a trigonal symmetry environment \cite{1987_GDavies_JPC, 1978_Tkachev_JAS, 1984_Burger_PRL}. Symmetry analysis revealed that the excited state may interact with higher lying doubly degenerate states \cite{1987_GDavies_JPC}.

A number of models were initially proposed to describe the W-center, including the $\langle111\rangle$-split-triple di-interstitial, tri-interstitial, and tetra-interstitial \cite{WC-theory1, WC-theory2, WC-theory3, WC-theory4}. The most energetically stable of these, the tri-interstitial I$_3$-I possessed a C$_{3v}$ symmetry and corresponding symmetric vibration mode which matched well with experiment \cite{WC-theory1, WC-theory2}. An alternate structure, I$_3$-II, was also proposed \cite{WC-theory-II}, and although it had a 1.4 eV lower formation energy, the tetrahedral symmetry did not allow for a local vibrational mode (LVM) compatible with the 70 meV sideband seen in the experimental PL spectrum. A third defect structure, dubbed I$_3$-V, with a formation energy lying between that of the previous two candidates \cite{WC-theory-V1, WC-theory-V2}, was discovered using molecular dynamics simulations \cite{WC-theory-Vdiscov}. While this configuration had the appropriate symmetry and exhibited a LVM with the correct energy \cite{WC-theory-V1,WC-theory-V2}, the electronic structure of its defect levels \cite{WC-theory-V1,WC-theory-V2,WCsingle} is still hotly debated, and it has been proposed that the optical transition stems from the recombination of an exciton localized at the defect by Coulomb interactions, even in the absence of defect levels in the band gap \cite{WCsingle}.

Another such luminescent defect in silicon is the T-center. It is less well studied than other defect centers in silicon \cite{1989_GDavies_PR}, but is attracting more interest recently due to its ZPL transition of 935 meV, which lies directly in the telecommunications $O$-band. Furthermore, this defect center is believed to host transitions between two spin-1/2 states whose fine structure can be controlled by an external magnetic field, presenting the possibility of using the T-center for photon-spin coupling for QIS applications \cite{tcenter-28Si}.

The T-center transition was identified as a doublet pair of states split by 1.75 meV, with uniaxial stress measurements indicating non-linear energy dependence of the defect levels \cite{tcenter-safonov, tcenter-irion}. The transition is thought to be between a highly isotropic spin-$1/2$ level and a highly anisotropic spin-$1/2$ level. Specifically, the ground state possesses a single unpaired electron occupying a mid-gap state, and when transitioning to the excited state, an additional bound exciton is created. The two bound electrons form an $S=0$ singlet within the mid-gap, while the $j=3/2$ hole state splits into two doublets, consistent with the experimentally observed peaks in the PL.

Perhaps the most extensively studied defect centers in silicon is the G-center, which consists of two carbon atoms and one silicon atom. The G-center forms when a substitutional carbon C$_{(s)}$ combines with an interstitial carbon C$_{(i)}$. C$_{(i)}$ defects are relatively mobile at 300 K and can be formed by radiation damage displacing another C$_{(s)}$ or by direct injection \cite{1989_GDavies_PR}. The G-center emission peak is located at 969 meV, with a relatively narrow linewidth of a few meV depending on synthesis conditions \cite{GCensemble, defect-engineering,qubit-exploration}. 

Uniaxial stress perturbation measurements were used to identify the symmetry of the G-center as C$_{1h}$, with the mirror plane along the $\langle 110 \rangle$ crystalographic axis. Early on it was suggested that there were two possible structures that satisfied this symmetry \cite{1989_GDavies_PR, GC1, GCensemble}, the Type-A (GCA) configuration where the defect atoms form a bent 
C$_{(s)}$-Si$_{(s)}$-C$_{(i)}$ chain with the carbon atom in the interstitial position (Fig \ref{struct}a), and the Type-B (GCB) configuration where the silicon atom shifts to the interstitial position, making a C$_{(s)}$-Si$_{(i)}$-C$_{(s)}$ chain (Figure \ref{struct}b). While under $n$-type doping conditions, switching between the two configurations was observed \cite{GC1}, subsequent works point to the G-center Type-B configuration being the true ground state \cite{GCensemble, gali-GC}. Optically detected magnetic resonance (ODMR) experiments have also been performed for the G-center, revealing a transient excited spin triplet state for which a sizable ZFS was observed \cite{zfs-exp}. 

Despite extensive study, the electronic structure (and to a lesser extent the atomic structure) of the G-center are still not settled. Early cluster calculations focused on understanding the relative stability of the Type-A and Type-B G-centers when positively or negatively charged, and revealed two defect levels within the silicon band gap corresponding to the two charge states \cite{ch-defects,GC2}. Later calculations of G-center defects embedded in a periodic supercell \cite{GC3,Gcenter-HSE} produced only a single defect level within the gap for the neutral defect state. Most recently, a combined approach of hybrid functionals and an additional Hubbard-$U$ parameter of $U=7.3 eV$ extracted from a GW calculation was used to reproduce the two defect levels at the $\Gamma$-point, one of which was located just below the valence band maximum \cite{gali-GC}.

\section{II. Electronic Structure of the Defect Centers}

In order to compute the electronic structures of these three defect centers, first principles density functional theory (DFT) calculations are performed with the Vienna \textit{ab initio} simulation package (VASP) \cite{vasp1,vasp2,vasp3,vasp4}, using the Heyd-Scuseria-Ernzerhof (HSE06) \cite{hse,Gcenter-HSE} functional. The G-center Type A (GCA) and Type B \cite{GC1,GC2,GC3,Gcenter-HSE} (GCB), W-center Type-V \cite{WC-theory-Vdiscov, WC-theory-V1, WC-theory-V2} (WCV), and T-center \cite{tcenter-safonov} (TC) defect structures were embedded within a $3\times 3\times 3$ supercell of silicon containing 216 silicon atoms, and relaxed to a force tolerance of 0.001 (eV/\AA) on a $2\times 2 \times 2$ $\Gamma$-centered $k$-point grid. The resulting structures and charge densities were used to compute the various defect properties including ZPL, ZFS, and orbital projections on the same $2\times 2 \times 2$ $k$-point grid aside from the zero-field splitting which was computed at the $\Gamma$-point. An energy cutoff of 450 eV was used. The relaxation procedure was also performed on a reduced $2\times 2\times 2$ supercell containing 64 atoms, and the obtained self-consistent charge density was then used to compute the band structure of each defect. Performing the full band structure calculation shows the dispersion throughout the Brillouin zone, revealing defect levels that would be hidden within the conduction or valence bands when doing calculations using only the $\Gamma$-point. This is in fact the case for both TC and GCA/GCB, where the lower defect state dips into the conduction band at the $\Gamma$-point. All of the calculations were performed in the neutral charge state of the defects, as there is no experimental evidence that their PL transitions are accompanied by a change in charge state.

Excited state calculations for each defect are preformed by manually constraining the orbital occupations to excite one electron into an unoccupied band. Such an approach has been successfully used in combination with hybrid functionals to study the excited state properties of the nitrogen-vacancy (NV) defect in diamond \cite{gali-constrainedNV}, as well as various silicon defects \cite{sinead-TC,gali-GC,WCsingle}. 

The computed band structures for the color centers provide additional information about which bands correspond to the local defect state at each $k$-point, provided that the defect states are well localized. However, if the defect states are not well localized, it is not possible to determine where they lie in the band order from an atom-projected band structure plot. The information contained in the projected band structure plots can be used to improve the accuracy of the constrained occupation method, by using the positions of the defect levels in the band order of the ground state to perform excited state calculations of the bands structure on the same $k$-path, as well as the excited state properties on a $2 \times 2 \times 2$ $k$-point grid. For the excited state calculations, the electron occupations are constrained independently at each $k$-point by removing an electron from the band corresponding to the lower defect level, and adding an electron to the band corresponding to the upper defect level. As the band order, and hence the positions of lower/upper defect levels, changes at different $k$-points, the constraints must be set independently for each $k$-point. This procedure is performed for TC, WC, and the optically active GCB, in conjunction with an additional structural relaxation in the constrained excited state to obtain the true excited state energy. We note that there is a functional limitation to this approach --- setting the initial occupations places electrons into states that do not correspond to a converged self-consistent charge density in the excited state. In particular, occupying a state changes its energy, so when the self-consistency in the charge density is reached, the character of the occupied state may change, so the intended occupations might no longer correspond to the local defect levels.

%putting T-center figure here so that it ends up on 2nd page
\begin{figure}[t]
	\includegraphics[width=1.0\columnwidth]{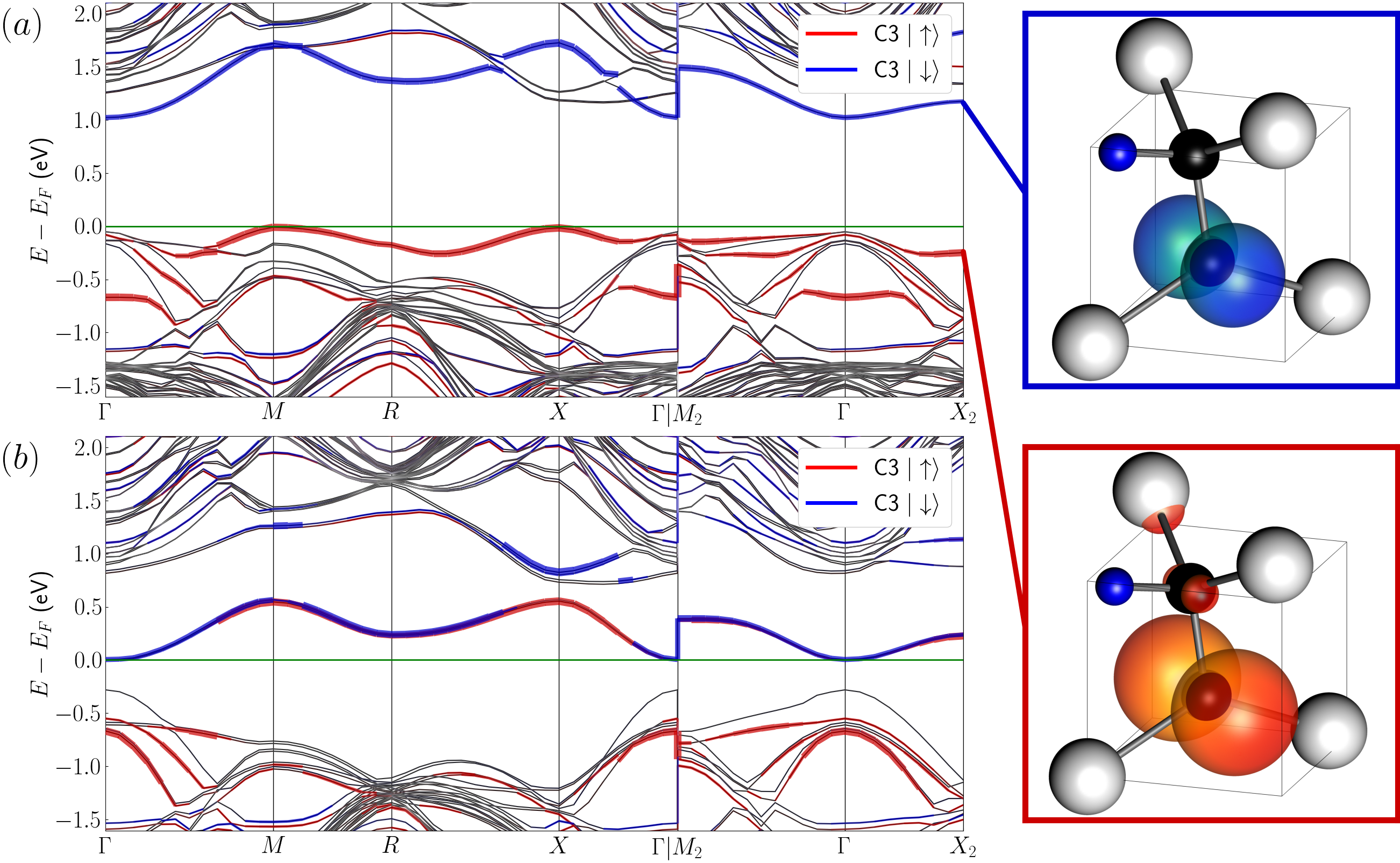}
	\caption[Band structures and orbitals of T-center]{Band structures of the (a) ground state and (b) excited state of the T-center, with spin up/down defect states highlighted in red/blue. The orbitals of the localized defect levels are shown on the right.}
	\label{tcenter}
\end{figure}

%TODO move (T-center)
The hybrid functional calculation of the silicon T-center produces the band structure shown in Figure \ref{tcenter}. In the ground state, the T-center possesses two defect levels that are coincident with the valence and conduction bands. Projecting onto the local orbitals reveals that both defect levels are highly localized to the trigonally bonded carbon atom, having the character of $p_z$ orbitals oriented perpendicularly to the $\langle 110 \rangle$ plane. The excited state band structure was also obtained, yielding a pair of degenerate defect levels directly in the middle of the silicon gap (Fig \ref{tcenter}b). This midgap level is fully occupied, consistent with the $S=0$ state predicted by experiments, and the defect bands retain the same $p_z$ character as the ground state. The spin moment of the T-center persists in the excited state, but is now associated with the hole state remaining in the valence band, resulting in the moment becoming delocalized over the entire supercell. Unfortunately, treating the electrons using constrained occupations does not give a complete description of the electron-hole interaction of the bound exciton, and hence cannot reproduce the $j=3/2$ configuration of the hole state.

%TODO move (W-center)
For the W-center, the band structure calculation produces two defect levels within the band gap (See Supplemental Material at \cite{supplement}), which are positioned near the conduction and valence band edges. Projection onto atomic orbitals reveals that both defect levels exhibit a low degree of localization -- the lower defect level is somewhat localized to the ring of silicon atoms formed by the three Si interstitials and the three Si atoms they displace, while the upper defect level is completely delocalized.

\begin{figure}[b]
	%\captionsetup{justification=raggedright}
	\includegraphics[width=1.0\columnwidth]{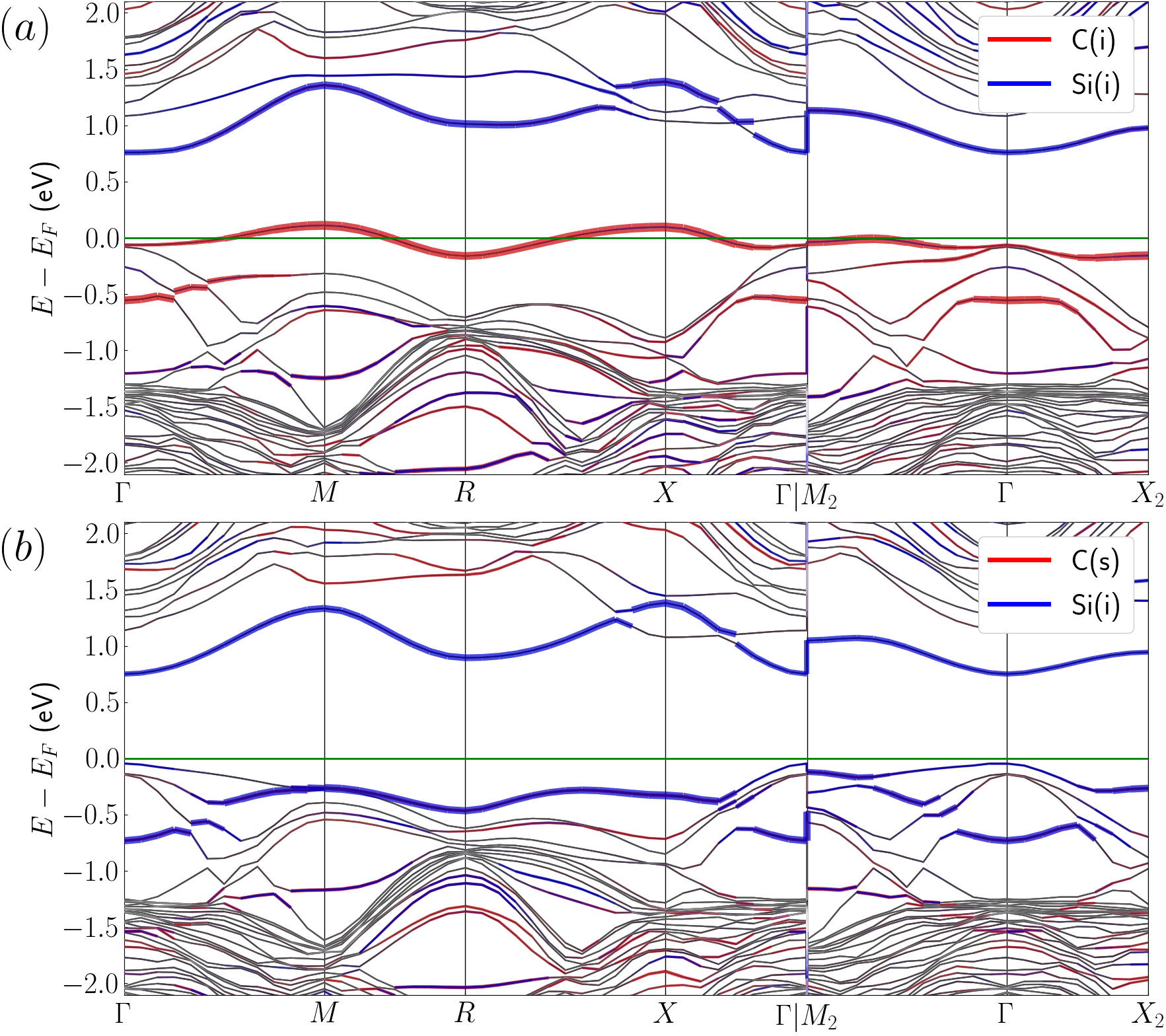}
	\caption[Band structures of G-center Type A and Type B]{Atom projected band structure of the silicon G-center Type A (a), and Type B (b) configurations. Contributions to the bands from the interstitial carbon atom (red) and interstitial silicon atom (blue) are highlighted. }
	\label{bands}
\end{figure}

%TODO Move (GC)
For the G-center, band structure calculations were performed for both configurations, GCA and GCB, in order to reveal the differences in their electronic structures. Figure \ref{bands} shows the band structures for GCA and GCB with the orbital projections onto the interstitial carbon and silicon atom of the defect highlighted in red and blue respectively. The band structure plots show that the two local defect levels intersect the valence and conduction bands. The upper defect level is resonant with the conduction band near the $X$-point, while the lower defect dips below the valence band maximum around the $\Gamma$-point, consistent with prior calculations on $\Gamma$-point only $k$-grids \cite{GC3,Gcenter-HSE,gali-GC}. It should be noted that although these defects are buried within the valence and conduction bands, local transitions between them can still be responsible for the observed photoluminescence. 

The plots also reveal a marked difference between the electronic structures of GCA and GCB. While the defect states for both centers are similarly well localized, the states belong to different sets of atoms. In GCA the defect levels are localized on separate atoms, with the lower energy state being localized to the interstitial carbon C$_{(i)}$, and the higher energy state being localized to the the substitutional silicon Si$_{(s)}$. In contrast, both defect levels in GCB are localized on Si$_{(i)}$. This results in increased interactions between electrons occupying the two-electron defect state, leading to an enhancement of the ZFS, which will be discussed in a subsequent section.

The localization of the defect orbitals additionally has a dramatic impact on the optical properties of the G-center. In fact the experimental PL spectrum of the G-center is attributed to the GCB structure, as GCA is considered to be optically inactive \cite{GC1, 1989_GDavies_PR}. The relative optical activity of the two structures can be compared by computing the electric dipole moments for the transitions between the two defect states at the $\Gamma$-point. For GCB, the value of the transition dipole moment is $\sim$4.3 Debye$^2$, while for GCA it is over two orders of magnitude smaller at $\sim$ 0.039 Debye$^2$, leading to significantly reduced optical activity.

\begin{table}[t]
    \caption{Values of the computed zero phonon line (ZPL) for G-center Type B (GC), T-center (TC), and W-center type V (WCV), which are compared with the experimental values. }
    \vspace{0.5cm}
	\begin{tabular}{ c c c }
		\hline
		\hline
		Defect & ZPL (meV) & ZPL (meV)\\
		Defect & Theory & Expt.\\\hline
		GC & 987 & 969 \\
		TC & 921 & 935 \\
		WC & 1203 & 1018\\
		\hline
		\hline
	\end{tabular}
	\label{zpldat}
\end{table}

The ZPL energies for the defects are obtained using the constrained occupation approach on a $2 \times 2 \times 2$ $k$-point grid, by occupying the bands corresponding to the defect levels in the ground and excited states independently at each $k$-point using the band order determined from the band structure calculation. The computed ZPL energies are shown in Table \ref{zpldat}. For TC and GC, the defect states are well localized, making it possible to define occupations for the excited state. This results in computed ZPL values that are within 2\% of the experimental values for the T-center and G-center. ZPL of 987 meV for the G-center, computed with hybrid functionals, compares well with a recent work which obtained a value of 985 meV using a combined GW and HSE+U method. The G-center triplet excited state is slightly lower in energy, lying 591 meV above the ground state. For the W-center, the defect levels are not well localized, so the atom-projected band structure cannot be used to identify the bands corresponding to the defect at each $k$-point. Thus, the calculation is performed by constraining occupations to the lowest unoccupied band, at each $k$-point which results in a significant error in the value of the ZPL as compared to experiment.

\section{IV. Photoluminescence}

A key property of quantum defect centers relevant for QIS is their ability to produce indistinguishable photons. In order to investigate the optical properties of the G-center, W-center, and T-center, we compute the PL spectra using a procedure originally developed for the NV-center in diamond \cite{Alkauskas_2014}. The normalized luminescence intensity $I(\hbar \omega) = C \omega^3 A(\hbar \omega)$, can be computed from the optical spectral function

\begin{equation}
	A(\hbar \omega ) = \sum_m \left| \left\langle\chi_{gm} \vert \chi_{e0} \right\rangle \right|^2 \delta(E_{ZPL} - E_{gm} -\hbar\omega),
\end{equation}
where the first subscript ($g$ or $e$) denotes ground and excited state vibrational levels $\chi$, and the second index ($0$ or $m$) denotes the vibrational mode. Likewise $E_{gm}$ is the energy of vibrational mode $m$ of the ground state, while $E_{ZPL}$ is the ZPL energy. The prefactor $C$ is a normalization constant, given by $C = \int A(\hbar \omega) \omega^3 d(\hbar \omega).$ Due to the difficulty of computing the overlap integrals $\left\langle\chi_{gm} \vert \chi_{e0} \right\rangle$, we opt instead to use the generating function approach where the optical spectral function is derived from a generating function $G(t)$ of the electron-phonon coupling spectral function $S(\hbar \omega)$:
\begin{equation}
	A(E_{ZPL} - \hbar \omega) = \frac{1}{2\pi} \int_{-\infty}^{\infty} G(t) e^{i\omega t - \gamma |t|} dt,
\end{equation}
where $\hbar\omega$ is the photon energy, $E_{ZPL}$ is the ZPL energy and $\gamma$ is the line broadening. The generating function is defined as 
\begin{equation}
	G(t) = e^{S(t) - S(0)},
\end{equation}
where $S(t)=\int_{0}^{\infty} S(\hbar \omega) e^{-i\omega t} d{\hbar \omega}$.
%\begin{equation}
%	S(t) = \int_{0}^{\infty} S(\hbar \omega) e^{-i\omega t} d{\hbar \omega}.
%\end{equation}
The electron-phonon spectral function can in turn be computed as a sum over the phonon modes $\lambda$ and wave vectors {\bf q}:
\begin{equation}
	S(\hbar \omega) = \sum_{{\bf q}\in {\rm BZ}}\sum_\lambda \frac{\omega_\lambda({\bf q}) f_\lambda({\bf q})^2}{2\hbar} \delta(\hbar \omega - \hbar \omega_\lambda({\bf q}))\,,
\end{equation}
where $\hbar\omega_\lambda({\bf q})$ is the energy of the phonon mode $(\lambda,{\bf q})$ and the coefficients $f_\lambda({\bf q})$ are given by:
\begin{equation}
	f_\lambda({\bf q}) = \sum_{\alpha i} m_\alpha^{1/2} \left(R_{e,\alpha i} - R_{g,\alpha i}\right) \epsilon_{\lambda,\alpha i}({\bf q})\,,
\end{equation}
where $\epsilon_{\lambda,\alpha i}({\bf q})$ is the coefficient of the phonon eigenvector. Further details on the calculation are provided in the supplementary material \cite{supplement}.

\begin{figure}%[hb]
	%\captionsetup{justification=raggedright}
	\includegraphics[width=1.0\columnwidth]{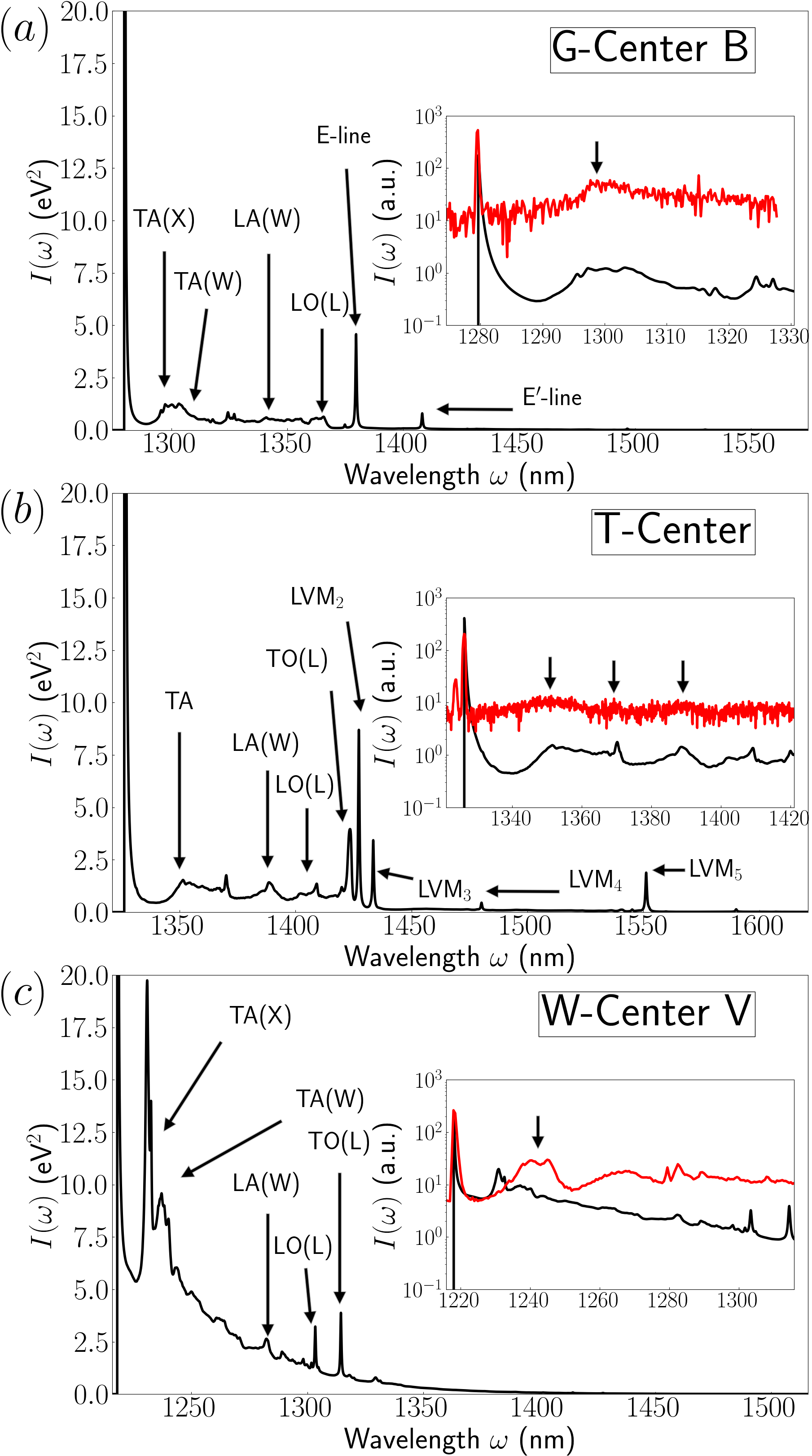}
	\caption[Photoluminescence of G, T and W centers.]{Computed photoluminescence spectra for (a) G-center Type B, (b) T-center, and (c) W-center Type V, adjusted to the experimental ZPL. Arrows indicate peaks in the sideband arising from the bulk silicon structure and local defect modes. Inset plots compare the computed (black) and experimental (red) PL spectra, with arrows indicating features of interest in the sideband.}
	\label{pl}
\end{figure}

The computed PL spectra for the G-center, T-center, and W-center are shown in Figure \ref{pl}. Only the GCB PL spectrum is shown since the GCA spectrum is very similar in appearance. The theoretical calculation is able to capture several characteristic features of the PL spectra. Each PL spectrum features a strong excitation peak which is aligned to coincide with the experimental ZPL. The $\sim 100$ nm above the ZPL peak feature a sideband stemming from the bulk silicon structure, and several standard peaks can be identified \cite{GCensemble}, including the transverse acoustic (TA), transverse optical (TO), longitudinal optical (LO), and longitudinal acoustic (LA) phonons at the $X$, $W$, and $L$ high-symmetry points of the Brillouin zone. Normally, such single-phonon peaks at these points would not appear in the PL spectrum due to momentum conservation, but in here they appear as a result of defect mediated processes which break the translation symmetry of the crystal. A number of LVM peaks for the G-center \cite{GCensemble, gcenter-eline} (E and E$^\prime$ lines), and T-center \cite{tcenter-lvm} are identified in the PL.

In particular, there are several features in the PL spectra that stand out. The W-center PL has a broad shoulder that is absent from the PL spectra of the other two defects, along with a lack of clear LVM peaks, which suggests a strong hybridization between the W-center local modes and silicon TA modes may be responsible. In fact, in the relaxed atomic structure of the excited state, the atoms are significantly displaced from their ground state positions, resulting in a broadening of the spectral function \cite{supplement} and PL spectrum. The TA peak in the T-center PL (Fig. \ref{pl} b), is quite broad, making it difficult to identify the contributions from the $X$ and $W$ points in the Brillouin zone, which usually form a peak and shoulder, as seen in the W-center PL (Fig. \ref{pl} c). Another reason for this difficulty is a small peak at the right edge, which might be attributed to the LVM$_1$ \cite{tcenter-lvm}. On the other hand, for the G-center, the prominent TO peak is essentially absent, consistent with experimental measurements \cite{GCensemble}. Furthermore, the LVM $E^\prime$-line \cite{gcenter-eline} is found much closer to the ZPL than in experiment, where it appears around $\sim 1500$ nm. A more detailed analysis of these features would necessitate the calculation of phonon-resolved PL spectra, which are beyond the scope of this work.

The computed spectra are compared with the experimentally obtained PL in the insets of Fig. \ref{pl}. Details on the experimental conditions are provided in the Supplemental Material \cite{supplement}. To facilitate the comparison, the theory spectra are aligned with the ZPL of the experimental data. There is fairly good agreement between theory and experiment, with the theoretical calculation capturing the main phonon peaks in the sideband. This is particularly evident for the GC, where the TA phonon peak at $\sim$1300 nm is nearly exactly matched by the calculation. For the WC, the TA peak at $\sim$1240 nm is also reproduced, although it appears at a slightly higher wavelength in experiment. For the T-center, the sideband peaks are quite broad, but nevertheless the bulk-silicon phonon peaks at $\sim$1350 nm, $\sim$1370 nm and $\sim$1390 nm line up well with experiment \cite{tcenter-SOI}. It is also important to highlight that the experimental PL of the G-centers in SOI has a remarkably narrow ZPL linewidth of $\sim 0.17$nm in this example, significantly narrower than the recently reported ZPL of 1.1 nm for a G-center in a waveguide \cite{gcenter-mit-waveguide} but slightly broader than the 0.1 nm linewidth observed for G-centers near the surface of bulk silicon samples \cite{qubit-exploration}. Strain in the SOI device layer is the likely cause for this broadening.

\begin{table}%[h]
    \caption[Huang-Rhys factors G-center, T-center and W-center.]{Huang-Rhys and Debye-Waller factors for G-center, T-center and W-center computed using hybrid functionals.}
    \vspace{0.5cm}
	\begin{tabular}{ c c c }
		\hline
		\hline
		Defect & Huang-Rhys factor & Debye–Waller factor \\
		\hline
		GCA & 0.266 & 0.766\\
		GCB	& 0.187 & 0.829\\
		TC  & 0.155 & 0.856\\
		WCV & 0.536 & 0.585\\
		\hline
		\hline
	\end{tabular}
	\label{hr}
\end{table}

The Huang-Rhys factors (HR) for each defect were also computed, and are shown in Table \ref{hr}. The Debye-Waller factor $w_{\text{DB}}$ is obtained from the Huang-Rhys factor $\bar{S}$ through the relation $w_{\text{DB}} = e^{-\bar{S}}$ There is a clear enhancement of the Huang-Rhys factor of the W-center as compared to the other defects, which stems from the highly delocalized nature of its defect states, which would couple more strongly to the bulk silicon phonon modes. This is consistent with the PL spectrum of the W-center which has a broad shoulder due to the strong hybridization of the local modes and silicon TA modes. Our finding that the GCB has a weaker electron-phonon coupling than the GCA can be explained by the GCB having defect states localized on a single atom, the silicon interstitial Si$_(i)$, compared to the GCA states being on separate atoms, C$_(i)$ and Si$_(s)$, and hence more susceptible to bond length fluctuations.

\section{V. Zero-Field Splitting}

Quantum defects hosting an optically accessible triplet state can serve as platforms for building quantum devices with coupled spin-photon degrees of freedom \cite{tcenter-quantum-computing,tcenter-spin-photon,qubit-diamond, qubit1,qubit2,qubit3,qubit4}. Although no triplet state has been observed for the W-center, there is some evidence for mixing between the excited state and some higher-lying doubly degenerate states \cite{1989_GDavies_PR}.  The T-center transitions and their potential for a spin-photon interface has been discussed in \cite{tcenter-quantum-computing, tcenter-spin-photon}.  In the remainder of this section we focus on the excited triplet state of the G-center and its ZFS.  The ZFS is computed for a hypothetical configuration with equal spins occupying the lower and higher energy defect levels using a constrained occupation approach. The resulting ZFS is quite small ($D_{\text{avg}} = 0.8$ MHz), which is expected given the delocalized nature of the defect states. 

It has been shown that the silicon G-center can be driven into a metastable excited triplet state with spin-1 angular momentum \cite{zfs-exp}. Symmetry breaking would split such a triplet state into a state with magnetic quantum number $m_s=0$ and degenerate states with $m_s = \pm 1$. A number of schemes exist for encoding a qubit using these states. One such scheme uses the $m_s = \pm 1$ states to encode $|0\rangle$ and $|1\rangle$ qubit states and using the $m_s=0$ state as an ancillary level for manipulating the geometric phase \cite{qubit1,qubit2}. Another possibility is to construct the qubit from the triplet $m_s=0$ and a separate singlet $S=0,m_s=0$ state \cite{qubit3,qubit4}, which gives the additional advantage of decoupling the qubit from external magnetic fields, as both states have zero spin quantum number.

\begin{table}[t]
    \caption{The computed zero field splitting (ZFS) values for the excited spin triplet state of the G-center are given for Type A \& B structures, and compared with experiment. }
    \vspace{0.5cm}
	\begin{tabular}{ c c c c }
		\hline
		\hline
		& \multicolumn{3}{c}{ZFS (MHz)}	  \\
		Defect & $|D_{xx}|$ & $|D_{yy}|$ & $|D_{zz}|$ \\\hline
		GCA (th.)	& 166	& 210	& 376	\\
		GCB (th.)	& 152	& 964	& 1116	\\
		GC (th., \cite{gali-GC}) & 307 & 911 & 1218\\
		GC (exp.,\cite{zfs-exp})	& 142	& 800	& 941	\\
		\hline
		\hline
	\end{tabular}
	\label{zfsdat}
\end{table}

\begin{figure}[bp]
	\includegraphics[width=1.0\columnwidth]{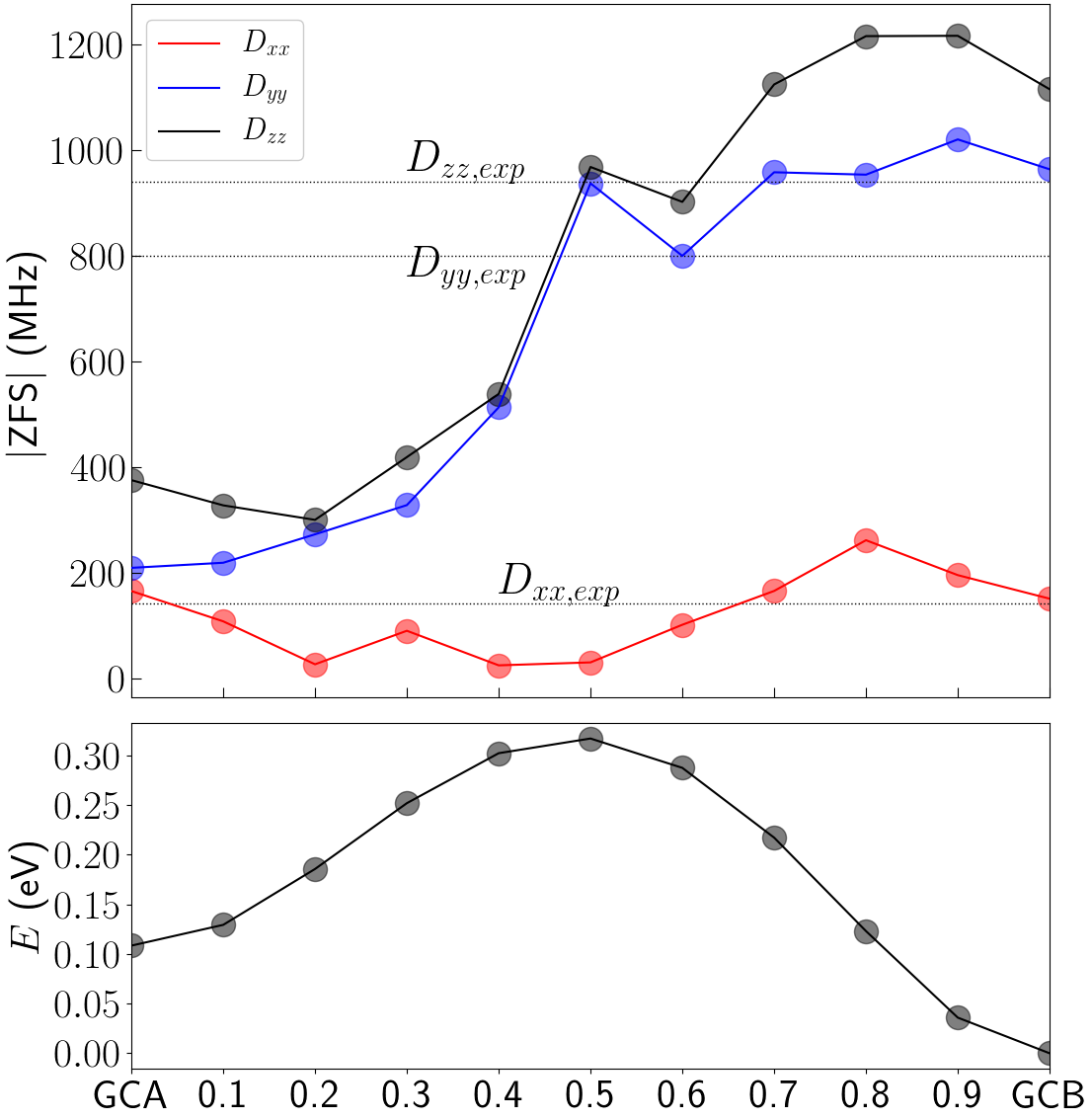}
 	\caption[ZFS and energy of the G-center, interpolating between GCA and GCB]{Zero field splitting (above) and defect energy (below) of the G-center for defect configurations that linearly interpolate between GCA and GCB in steps of 0.1 (10\%). Values of the diagonalized ZFS tensor are shown, with $D_{zz}$ in black, $D_{yy}$ in blue, and $D_{xx}$ in red. Dashed lines indicate the experimentally measured ZFS values.}
	\label{zfs}
\end{figure}

The fidelity of these schemes for encoding a qubit using triplet states is highly dependent on the separation of the $m_s=0$ and $m_s=\pm 1$ levels. In the absence of magnetic fields, this splitting is driven by spin-spin interactions of the electrons occupying the defect states. The spin-spin interaction can be written \cite{NV-group-theory}:

\begin{equation}
H_{ss} = \bm{S}\bm{D}\bm{S} = -\frac{\mu_0}{4\pi} \frac{g^2 \mu_B^2}{r^3} \left( \frac{3}{4} (\bm{s}_1\cdot \hat{\bm{r}})(\bm{s}_2\cdot \hat{\bm{r}}) - \bm{s}_1 \cdot \bm{s}_2 \right),
\end{equation}
where $\mu_0$ is the magnetic permeability, $g$ is the Land\'{e} g-factor, $\mu_B$ is the Bohr magneton, $\bm{D}$ is the traceless dipole-dipole interaction tensor, the total spin $\bm{S} = \bm{s}_1 + \bm{s}_2$, $\bm{s}_i = \frac{1}{2}[\sigma_x,\sigma_y,\sigma_z]$ is the spin operator for particle $i$, and $\sigma_x,\sigma_y,\sigma_z$ are the Pauli matrices. For the excited triplet state of the G-center, $\psi$, this spin-spin interaction leads to a gap $3\Delta$ between the $m_s=0$ and $m_s=\pm 1$ where

\begin{equation}
	3\Delta = -\frac{3}{4} D_{zz} = 3 \frac{\mu_0}{4\pi} g^2 \mu_B^2 \left\langle \psi \left| \frac{1-3\hat{z}^2}{4r^3} \right| \psi \right\rangle.
	\label{zfsmain}
\end{equation}

The ZFS tensor was computed using hybrid functionals for both GCA and GCB in the excited triplet state. The absolute values of the diagonalized tensor are shown in Table \ref{zfsdat}, with the GCB values showing good agreement with experiment \cite{zfs-exp}. Note that the ZPL is traceless, i.e.  the values on the diagonal must satisfy the condition $D_{xx} + D_{yy} + D_{zz} = 0$, and are automatically ordered by magnitude $D_{zz} > D_{yy} > D_{xx}$. Similar computed values for the G-center were recently reported \cite{gali-GC}. We point out that the method for simulating the G-center used in \cite{gali-GC} requires a Hubbard-$U$ value of 7.3 eV, when even values of 0.4 eV are considered large for silicon defects \cite{tcenter-safonov}. 
Additionally, the inclusion of a Hubbard-U parameter only serves to shift the energies of the defect levels, and does not significantly affect the localization.
As the subsequent discussion will demonstrate, the magnitude of the ZFS is heavily dependent on the localization of the defect wavefunctions, and indeed the magnitude of ZFS computed by HSE and HSE+U is comparable \cite{gali-GC}.
In contrast, our approach only relies on hybrid functionals, which have been well-established for reproducing the electronic structure of semiconductors such as silicon. 

Both the calculation and experiment show a dramatic difference in the magnitudes of the ZFS of GCA and GCB. The evolution of the ZFS diagonal values is tracked as the G-center atomic positions are linearly interpolated between the two structures. The magnitudes of the ZFS for each structure along the interpolation are compared to the experimental values in Figure \ref{zfs}. It can be seen that while the $D_{xx}$ component remains relatively small, the $D_{yy}$ and $D_{zz}$ components experience a sudden step increase right at the intermediate point between the GCA and GCB structures.

\begin{figure}[hb]
	\includegraphics[width=1.0\columnwidth]{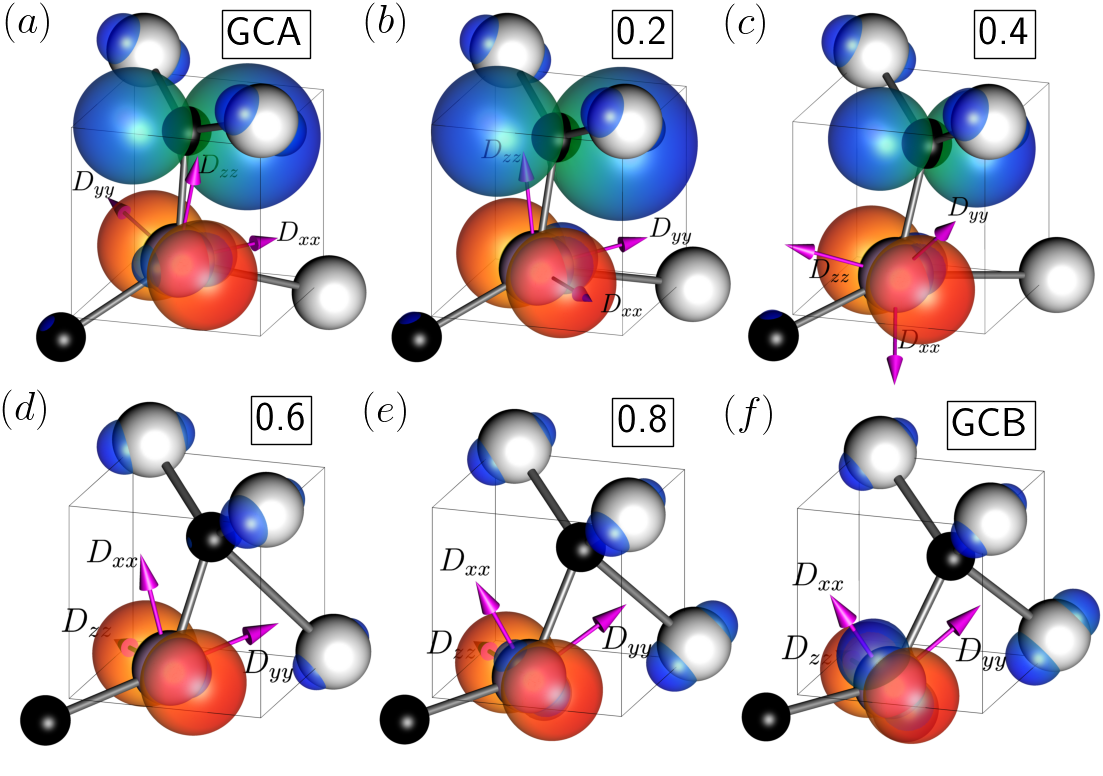}
	\caption[Orbital projections for G-center defect levels, interpolating between GCA and GCB]{Orbital projections of G-center lower (blue) and higher (red) energy defect levels for structures linearly interpolating between GCA and GCB in steps of 0.2. Arrows centered on the $Si(i)$ of the G-center defect indicate the direction of the ZFS tensor components.}
	\label{proj}
\end{figure}

To understand the evolution of the ZFS components as the G-center transitions from the A-type to B-type configuration, we project the two single-electron defect levels onto the $s$ and $p$ orbitals of the six atoms in the local cluster (Fig. \ref{proj}). For the A-type structure, the lower defect level is localized to the interstitial carbon and oriented in the $\langle 110\rangle$ direction, while the upper defect level is localized to the central silicon atom of the defect, having the character of a $p$ orbital oriented along $\langle \bar{1}10\rangle$. As the structure transitions between GCA and GCB, the $C_{(i)}$ and Si atoms move downward, with the carbon ultimately assuming a substitutional position in the lattice. Concurrently, the contribution of the carbon orbitals to the defect level weakens, disappearing entirely at the structural midpoint between GCA and GCB. As the defect approaches the GCB configuration, the character of the upper defect level remains largely unchanged, while the lower defect level becomes localized to the Si atom and acquires the character of a $p$ orbital oriented along the $Si_{(i)}$-$C_{(s)}$ bond.

During the linear interpolation between GCA and GCB configurations, the total energy of the defect changes considerably, as shown in Figure \ref{zfs}. The GCB structure is $\sim 100$meV lower in energy than the GCA structure, establishing it as the ground state configuration, consisted with prior work \cite{GCensemble, gali-GC}. There is additionally a potential barrier of $0.2$eV for converting between the two configurations, which is slightly larger than the experimentally determined value of $0.15$eV \cite{GC1}. However, our computed value may overestimate the true kinetic barrier, since we consider only a linear interpolation between GCA and GCB, and not a nudged-elastic band method which would sample all possible structural pathways to determine the lowest energy transformation. 

The evolution of the defect states can be understood in terms of a simple nearest neighbor tight binding model on the defect cluster (see supplementary material \cite{supplement})
\begin{equation}
    \mathcal{H} = \sum_{i} \epsilon_i c^\dagger_i c^{\phantom{\dagger}}_i + \sum_{\langle ij \rangle} t_{ij} c^\dagger_i c^{\phantom{\dagger}}_j,
\end{equation}
where $\epsilon_i$ is the energy of the orbital on the C or Si atom, $t_{ij}$ is the hopping energy between atoms, and the second sum runs over the nearest neighbors $\langle ij \rangle$. We can approximate all nearest neighbor hoppings as equal $t_{ij} \equiv t$, to demonstrate that the origin of the defect level structure is purely geometrical. In the G-center all of the atoms are tetrahedrally bonded to their nearest neighbors, except some of the defect atoms which have a lower coordination. For a tetrahedrally bonded atom, the natural basis is the set of $sp_3$ orbitals. Taking the energy $\epsilon_0$ of an unhybridized $p$ orbital as a reference point, let the energy of the four $sp_3$ orbitals be $\Delta\epsilon_1 = \epsilon_1 - \epsilon_0$, and the energy of the four orbitals on the nearest neighbor atoms be $\Delta\epsilon_2 = \epsilon_2 - \epsilon_0$. Diagonalizing the Hamiltonian of this cluster \cite{supplement} we find two sets of bands 
\begin{equation}
    E_{\text{bulk},\pm} = \frac{1}{2}\left(\Delta\epsilon_1 + \Delta\epsilon_2 \pm \sqrt{(\Delta\epsilon_1 - \Delta\epsilon_2)^2 + 4t^2}\right),
    \label{bulk}
\end{equation}
which when taken together with all the other tetrahedrally bonded silicon atoms in the bulk, form the valence and conduction bands. When the energies of the four $sp^3$ orbitals and the nearest-neighbor orbitals are similar, $t \gg (\Delta\epsilon_1 - \Delta\epsilon_2)$, and the expression above reduces to $E_{\text{bulk},\pm} \approx (\Delta\epsilon_1 + \Delta\epsilon_2)/2 \pm t$, yielding two bands separated by a gap $\sim 2t$.

The defect levels in the G-center arise from the defect atoms having lower than tetrahedral coordination. In GCA, the Si and C$_{(i)}$ defect atoms are both trigonally bonded, and so can be represented using a basis consisting of the in-plane $sp^2$ orbitals and non-bonding $p_z$ orbital, oriented to align with the local geometry of each atom. Taking the energy of the $p_z$ orbital to be $\epsilon_0 = 0$, the energy of the $sp^2$ to be $\Delta\epsilon_1$, and the energies of the nearest-neighbor orbitals as $\Delta\epsilon_2$, we diagonalize the Hamiltonian of the defect atom. Once again, we recover two sets of degenerate bands of the form Eq. \ref{bulk}, but now the non-bonding $p_z$ orbital forms a single defect level in the middle of the gap. Given there are two such trigonal defect atoms, we can expect two defect levels in the gap, having the character of $p_z$ orbitals aligned with their local orientation. This is in fact consistent with the orbital projections shown in Figure \ref{proj} for GCA.

In GCB, the bonding geometry is quite different, with both carbons assuming tetrahedrally bonded substitutional positions, and the defect Si atom bonding only to the two nearest carbon atoms. Due to the bent shape of the C$_{(s)}$-Si$_{(i)}$-C$_{(s)}$ chain, the same $sp^2 + p_z$ basis is appropriate here, however now only two of the $sp^2$ orbitals bond with nearest neighbor atoms. The resulting eigenenergies \cite{supplement} are again the set of gapped bands of the form of Eq. \ref{bulk}, but now with two defect levels in the gap, one for each of the non-bonding $sp^2$ and $p_z$ orbitals. The same result is found for the self-consistent simulation of GCB; Figure \ref{proj} shows the projections of the two defect levels match the two orbitals derived using a nearest-neighbor model on the defect cluster.

The argument for why the ZFS is enhanced in GCB can be made more explicit by comparing the ZFS integrals given by Eq. \ref{zfsmain} for the approximate two-electron defect states previously derived for GCA and GCB \cite{zfs-derivation}. It can be shown that for orbitals localized to different atoms of the defect, separated by vector $\bm{d} = (d,0,0)$, the main contribution to the ZFS (Eq. \ref{zfsmain}) will be of the form $C/d^3$, where $C$ is a factor arising from integration over internal variables. For single electron states which are well localized to their separate defect atoms, as is the case for GCA, $D_{zz}$ will approach this limiting form, and scale inversely with $1/d^3$, resulting in a decreased magnitude of the ZFS. For GCB, the two-electron state is composed of orbitals located on the same atom, and can therefore achieve a much larger value of ZFS.

%The following is probably incorrect, better to remove it
%Since the ZFS is essentially a measure of the anisotropy of the two-particle electronic density in the triplet state, the orientation of the $D$ tensor diagonal elements can be understood over the course of the transition. Initially, the orbitals comprising the two-electron state are localized to different atoms, for GCA, 0.2, and 0.4 fraction configurations, resulting in a small magnitude of the ZFS overall. The orientation of the ZFS components then becomes set by small random anisotropy, so the largest ($D_{zz}$) and next-largest ($D_{yy}$) components become oriented randomly. For the 0.6, 0.8, and GCB configurations, the two orbitals comprising the many body state hosting the spins in the triplet configuration both become localized to the silicon atom, and the ZFS tensor components align with the orbital orientations. $D_{zz}$, the largest component, becomes aligned with the higher energy level (red) in the $\langle -110\rangle$ direction, which is the higher probability projection of the two single-electron defect states. The second largest component, $D_{yy}$, aligns with the orbital corresponding to the lower energy single-electron defect level, orienting in the $\langle 111\rangle$ direction.

\section{VI. Conclusion}

In this work, we have presented an overview of three different defect centers in silicon that have potential QIS applications, the G-center, T-center, and W-center. Using a hybrid functional approach, we have computed the electronic structure ZPL, ZFS, and PL spectra for these defects. In particular, we have shown how the physical structure impacts the electronic structure of the G-center, resulting in a significant enhancement of the ZFS in the Type-B configuration. Specifically, we investigate the zero-field splitting for the excited triplet state of the G-center defect as the structure is linearly interpolated from the A-configuration to the B-configuration, showing a sudden increase in the magnitude of the $D_{zz}$ component of the zero-field splitting tensor. This transition is explained in terms of localization of the G-center defect states onto the Si interstitial. 

We have also shown how the excited state associated with the ZPL transition in the T-center has a complete delocalization of the magnetic moment even though the defect states are well localized, which provides theoretical insight for engineering coupling between optical degrees of freedom and spin localization. Finally, we establish the delocalized nature of the defect levels in the W-center, and discuss how that leads to a stronger electron-phonon coupling, ultimately resulting in a broadened PL spectrum and an increased Huang-Rhys factor.

Finally, the tight-binding models derived in this work provide a simple framework for designing defects with bonding arrangements that favor the localization of both spins in a defect triplet state to the same atom, resulting in an improved ZFS while keeping HR factors low. Such models are quite robust, in the sense that they may be easily applied to other color centers in silicon or other host materials, enabling the prediction of a wide range of desirable defects. Future work will focus on finding such defects that are also ground-state triplets while being thermodynamically stable, setting up the possibility of creating them through novel methods such as those demonstrated recently to reliably create G- and W-centers \cite{defect-engineering}. Alternatively, excited-state triplets may provide transient access to a hyperfine-coupled nuclear spin for a practical spin-photon interface.

\section{Acknowledgements}

Experimental work was done by WL, KJ, WR, YZ, CP, and WQ. Samples were synthesized by WL, KJ, TS, AP and WQ, and PL measurements were performed by WL, WR, YZ, and CP. VI, JS, YL, and LZT performed the theoretical calculations. VI, JS, YL performed the numerical simulations, PL spectra were computed by JS and VI, tight-binding models done by VI. VI, JS, LT, and WL wrote the manuscript, edits by other authors. LT, TS, and BK supervised the project.

C. Papapanos' research for this paper was supported in part by Onassis Foundation Scholarship. This work was supported by the Office of Science, Office of Fusion Energy Sciences, of the U.S. Department of Energy, under Contract No. DE-AC02-05CH11231.  LZT, VI, YL and JS were also supported by the Molecular Foundry, a DOE Office of Science User Facility supported by the Office of Science of the U.S. Department of Energy under Contract No. DE-AC02-05CH11231.   This research used resources of the National Energy Research Scientific Computing Center, a DOE Office of Science User Facility supported by the Office of Science of the U.S. Department of Energy under Contract No. DE-AC02-05CH11231.

\vfill\null

\bibliography{references}

\begin{thebibliography}{53}%
\makeatletter
\providecommand \@ifxundefined [1]{%
 \@ifx{#1\undefined}
}%
\providecommand \@ifnum [1]{%
 \ifnum #1\expandafter \@firstoftwo
 \else \expandafter \@secondoftwo
 \fi
}%
\providecommand \@ifx [1]{%
 \ifx #1\expandafter \@firstoftwo
 \else \expandafter \@secondoftwo
 \fi
}%
\providecommand \natexlab [1]{#1}%
\providecommand \enquote  [1]{``#1''}%
\providecommand \bibnamefont  [1]{#1}%
\providecommand \bibfnamefont [1]{#1}%
\providecommand \citenamefont [1]{#1}%
\providecommand \href@noop [0]{\@secondoftwo}%
\providecommand \href [0]{\begingroup \@sanitize@url \@href}%
\providecommand \@href[1]{\@@startlink{#1}\@@href}%
\providecommand \@@href[1]{\endgroup#1\@@endlink}%
\providecommand \@sanitize@url [0]{\catcode `\\12\catcode `\$12\catcode
  `\&12\catcode `\#12\catcode `\^12\catcode `\_12\catcode `\%12\relax}%
\providecommand \@@startlink[1]{}%
\providecommand \@@endlink[0]{}%
\providecommand \url  [0]{\begingroup\@sanitize@url \@url }%
\providecommand \@url [1]{\endgroup\@href {#1}{\urlprefix }}%
\providecommand \urlprefix  [0]{URL }%
\providecommand \Eprint [0]{\href }%
\providecommand \doibase [0]{http://dx.doi.org/}%
\providecommand \selectlanguage [0]{\@gobble}%
\providecommand \bibinfo  [0]{\@secondoftwo}%
\providecommand \bibfield  [0]{\@secondoftwo}%
\providecommand \translation [1]{[#1]}%
\providecommand \BibitemOpen [0]{}%
\providecommand \bibitemStop [0]{}%
\providecommand \bibitemNoStop [0]{.\EOS\space}%
\providecommand \EOS [0]{\spacefactor3000\relax}%
\providecommand \BibitemShut  [1]{\csname bibitem#1\endcsname}%
\let\auto@bib@innerbib\@empty
%</preamble>
\bibitem [{\citenamefont {Bergeron}\ \emph
  {et~al.}(2020{\natexlab{a}})\citenamefont {Bergeron}, \citenamefont
  {Chartrand}, \citenamefont {Kurkjian}, \citenamefont {Morse}, \citenamefont
  {Riemann}, \citenamefont {Abrosimov}, \citenamefont {Becker}, \citenamefont
  {Pohl}, \citenamefont {Thewalt},\ and\ \citenamefont
  {Simmons}}]{tcenter-spin-photon}%
  \BibitemOpen
  \bibfield  {author} {\bibinfo {author} {\bibfnamefont {L.}~\bibnamefont
  {Bergeron}}, \bibinfo {author} {\bibfnamefont {C.}~\bibnamefont {Chartrand}},
  \bibinfo {author} {\bibfnamefont {A.~T.~K.}\ \bibnamefont {Kurkjian}},
  \bibinfo {author} {\bibfnamefont {K.~J.}\ \bibnamefont {Morse}}, \bibinfo
  {author} {\bibfnamefont {H.}~\bibnamefont {Riemann}}, \bibinfo {author}
  {\bibfnamefont {N.~V.}\ \bibnamefont {Abrosimov}}, \bibinfo {author}
  {\bibfnamefont {P.}~\bibnamefont {Becker}}, \bibinfo {author} {\bibfnamefont
  {H.-J.}\ \bibnamefont {Pohl}}, \bibinfo {author} {\bibfnamefont {M.~L.~W.}\
  \bibnamefont {Thewalt}}, \ and\ \bibinfo {author} {\bibfnamefont
  {S.}~\bibnamefont {Simmons}},\ }\href {\doibase 10.1103/PRXQuantum.1.020301}
  {\bibfield  {journal} {\bibinfo  {journal} {PRX Quantum}\ }\textbf {\bibinfo
  {volume} {1}},\ \bibinfo {pages} {020301} (\bibinfo {year}
  {2020}{\natexlab{a}})}\BibitemShut {NoStop}%
\bibitem [{\citenamefont {Udvarhelyi}\ \emph {et~al.}(2021)\citenamefont
  {Udvarhelyi}, \citenamefont {Somogyi}, \citenamefont {Thiering},\ and\
  \citenamefont {Gali}}]{gali-GC}%
  \BibitemOpen
  \bibfield  {author} {\bibinfo {author} {\bibfnamefont {P.}~\bibnamefont
  {Udvarhelyi}}, \bibinfo {author} {\bibfnamefont {B.}~\bibnamefont {Somogyi}},
  \bibinfo {author} {\bibfnamefont {G.~m.~H.}\ \bibnamefont {Thiering}}, \ and\
  \bibinfo {author} {\bibfnamefont {A.}~\bibnamefont {Gali}},\ }\href {\doibase
  10.1103/PhysRevLett.127.196402} {\bibfield  {journal} {\bibinfo  {journal}
  {Phys. Rev. Lett.}\ }\textbf {\bibinfo {volume} {127}},\ \bibinfo {pages}
  {196402} (\bibinfo {year} {2021})}\BibitemShut {NoStop}%
\bibitem [{\citenamefont {Tyryshkin}\ \emph {et~al.}(2006)\citenamefont
  {Tyryshkin}, \citenamefont {Morton}, \citenamefont {Benjamin}, \citenamefont
  {Ardavan}, \citenamefont {Briggs}, \citenamefont {Ager},\ and\ \citenamefont
  {Lyon}}]{Tyryshkin_2006}%
  \BibitemOpen
  \bibfield  {author} {\bibinfo {author} {\bibfnamefont {A.~M.}\ \bibnamefont
  {Tyryshkin}}, \bibinfo {author} {\bibfnamefont {J.~J.~L.}\ \bibnamefont
  {Morton}}, \bibinfo {author} {\bibfnamefont {S.~C.}\ \bibnamefont
  {Benjamin}}, \bibinfo {author} {\bibfnamefont {A.}~\bibnamefont {Ardavan}},
  \bibinfo {author} {\bibfnamefont {G.~A.~D.}\ \bibnamefont {Briggs}}, \bibinfo
  {author} {\bibfnamefont {J.~W.}\ \bibnamefont {Ager}}, \ and\ \bibinfo
  {author} {\bibfnamefont {S.~A.}\ \bibnamefont {Lyon}},\ }\href {\doibase
  10.1088/0953-8984/18/21/s06} {\bibfield  {journal} {\bibinfo  {journal}
  {Journal of Physics: Condensed Matter}\ }\textbf {\bibinfo {volume} {18}},\
  \bibinfo {pages} {S783} (\bibinfo {year} {2006})}\BibitemShut {NoStop}%
\bibitem [{\citenamefont {Redjem}\ \emph {et~al.}(2022)\citenamefont {Redjem},
  \citenamefont {Amsellem}, \citenamefont {Allen}, \citenamefont {Benndorf},
  \citenamefont {Bin}, \citenamefont {Bulanov}, \citenamefont {Esarey},
  \citenamefont {Feldman}, \citenamefont {Fernandez}, \citenamefont {Lopez},
  \citenamefont {Geulig}, \citenamefont {Geddes}, \citenamefont {Hijazi},
  \citenamefont {Ji}, \citenamefont {Ivanov}, \citenamefont {Kante},
  \citenamefont {Gonsalves}, \citenamefont {Meijer}, \citenamefont {Nakamura},
  \citenamefont {Persaud}, \citenamefont {Pong}, \citenamefont {Obst-Huebl},
  \citenamefont {Seidl}, \citenamefont {Simoni}, \citenamefont {Schroeder},
  \citenamefont {Steinke}, \citenamefont {Tan}, \citenamefont {Wunderlich},
  \citenamefont {Wynne},\ and\ \citenamefont {Schenkel}}]{defect-engineering}%
  \BibitemOpen
  \bibfield  {author} {\bibinfo {author} {\bibfnamefont {W.}~\bibnamefont
  {Redjem}}, \bibinfo {author} {\bibfnamefont {A.~J.}\ \bibnamefont
  {Amsellem}}, \bibinfo {author} {\bibfnamefont {F.~I.}\ \bibnamefont {Allen}},
  \bibinfo {author} {\bibfnamefont {G.}~\bibnamefont {Benndorf}}, \bibinfo
  {author} {\bibfnamefont {J.}~\bibnamefont {Bin}}, \bibinfo {author}
  {\bibfnamefont {S.}~\bibnamefont {Bulanov}}, \bibinfo {author} {\bibfnamefont
  {E.}~\bibnamefont {Esarey}}, \bibinfo {author} {\bibfnamefont {L.~C.}\
  \bibnamefont {Feldman}}, \bibinfo {author} {\bibfnamefont {J.~F.}\
  \bibnamefont {Fernandez}}, \bibinfo {author} {\bibfnamefont {J.~G.}\
  \bibnamefont {Lopez}}, \bibinfo {author} {\bibfnamefont {L.}~\bibnamefont
  {Geulig}}, \bibinfo {author} {\bibfnamefont {C.~R.}\ \bibnamefont {Geddes}},
  \bibinfo {author} {\bibfnamefont {H.}~\bibnamefont {Hijazi}}, \bibinfo
  {author} {\bibfnamefont {Q.}~\bibnamefont {Ji}}, \bibinfo {author}
  {\bibfnamefont {V.}~\bibnamefont {Ivanov}}, \bibinfo {author} {\bibfnamefont
  {B.}~\bibnamefont {Kante}}, \bibinfo {author} {\bibfnamefont
  {A.}~\bibnamefont {Gonsalves}}, \bibinfo {author} {\bibfnamefont
  {J.}~\bibnamefont {Meijer}}, \bibinfo {author} {\bibfnamefont
  {K.}~\bibnamefont {Nakamura}}, \bibinfo {author} {\bibfnamefont
  {A.}~\bibnamefont {Persaud}}, \bibinfo {author} {\bibfnamefont
  {I.}~\bibnamefont {Pong}}, \bibinfo {author} {\bibfnamefont {L.}~\bibnamefont
  {Obst-Huebl}}, \bibinfo {author} {\bibfnamefont {P.~A.}\ \bibnamefont
  {Seidl}}, \bibinfo {author} {\bibfnamefont {J.}~\bibnamefont {Simoni}},
  \bibinfo {author} {\bibfnamefont {C.}~\bibnamefont {Schroeder}}, \bibinfo
  {author} {\bibfnamefont {S.}~\bibnamefont {Steinke}}, \bibinfo {author}
  {\bibfnamefont {L.~Z.}\ \bibnamefont {Tan}}, \bibinfo {author} {\bibfnamefont
  {R.}~\bibnamefont {Wunderlich}}, \bibinfo {author} {\bibfnamefont
  {B.}~\bibnamefont {Wynne}}, \ and\ \bibinfo {author} {\bibfnamefont
  {T.}~\bibnamefont {Schenkel}},\ }\href {\doibase 10.48550/ARXIV.2203.13781}
  {\enquote {\bibinfo {title} {Defect engineering of silicon with ion pulses
  from laser acceleration},}\ } (\bibinfo {year} {2022}),\ \Eprint
  {http://arxiv.org/abs/2203.13781} {arXiv:2203.13781} \BibitemShut {NoStop}%
\bibitem [{\citenamefont {Awschalom}\ \emph {et~al.}(2018)\citenamefont
  {Awschalom}, \citenamefont {Hanson}, \citenamefont {Wrachtrup},\ and\
  \citenamefont {Zhou}}]{Awschalom2018}%
  \BibitemOpen
  \bibfield  {author} {\bibinfo {author} {\bibfnamefont {D.~D.}\ \bibnamefont
  {Awschalom}}, \bibinfo {author} {\bibfnamefont {R.}~\bibnamefont {Hanson}},
  \bibinfo {author} {\bibfnamefont {J.}~\bibnamefont {Wrachtrup}}, \ and\
  \bibinfo {author} {\bibfnamefont {B.~B.}\ \bibnamefont {Zhou}},\ }\href
  {\doibase 10.1038/s41566-018-0232-2} {\bibfield  {journal} {\bibinfo
  {journal} {Nature Photonics}\ }\textbf {\bibinfo {volume} {12}},\ \bibinfo
  {pages} {516} (\bibinfo {year} {2018})}\BibitemShut {NoStop}%
\bibitem [{\citenamefont {Khoury}\ and\ \citenamefont
  {Abbarchi}(2022)}]{QIS-review}%
  \BibitemOpen
  \bibfield  {author} {\bibinfo {author} {\bibfnamefont {M.}~\bibnamefont
  {Khoury}}\ and\ \bibinfo {author} {\bibfnamefont {M.}~\bibnamefont
  {Abbarchi}},\ }\href {\doibase 10.1063/5.0093822} {\bibfield  {journal}
  {\bibinfo  {journal} {Journal of Applied Physics}\ }\textbf {\bibinfo
  {volume} {131}},\ \bibinfo {pages} {200901} (\bibinfo {year} {2022})},\
  \Eprint {http://arxiv.org/abs/https://doi.org/10.1063/5.0093822}
  {https://doi.org/10.1063/5.0093822} \BibitemShut {NoStop}%
\bibitem [{\citenamefont {Durand}\ \emph {et~al.}(2021)\citenamefont {Durand},
  \citenamefont {Baron}, \citenamefont {Redjem}, \citenamefont {Herzig},
  \citenamefont {Benali}, \citenamefont {Pezzagna}, \citenamefont {Meijer},
  \citenamefont {Kuznetsov}, \citenamefont {G\'erard}, \citenamefont
  {Robert-Philip}, \citenamefont {Abbarchi}, \citenamefont {Jacques},
  \citenamefont {Cassabois},\ and\ \citenamefont {Dr\'eau}}]{intro-Durand}%
  \BibitemOpen
  \bibfield  {author} {\bibinfo {author} {\bibfnamefont {A.}~\bibnamefont
  {Durand}}, \bibinfo {author} {\bibfnamefont {Y.}~\bibnamefont {Baron}},
  \bibinfo {author} {\bibfnamefont {W.}~\bibnamefont {Redjem}}, \bibinfo
  {author} {\bibfnamefont {T.}~\bibnamefont {Herzig}}, \bibinfo {author}
  {\bibfnamefont {A.}~\bibnamefont {Benali}}, \bibinfo {author} {\bibfnamefont
  {S.}~\bibnamefont {Pezzagna}}, \bibinfo {author} {\bibfnamefont
  {J.}~\bibnamefont {Meijer}}, \bibinfo {author} {\bibfnamefont {A.~Y.}\
  \bibnamefont {Kuznetsov}}, \bibinfo {author} {\bibfnamefont {J.-M.}\
  \bibnamefont {G\'erard}}, \bibinfo {author} {\bibfnamefont {I.}~\bibnamefont
  {Robert-Philip}}, \bibinfo {author} {\bibfnamefont {M.}~\bibnamefont
  {Abbarchi}}, \bibinfo {author} {\bibfnamefont {V.}~\bibnamefont {Jacques}},
  \bibinfo {author} {\bibfnamefont {G.}~\bibnamefont {Cassabois}}, \ and\
  \bibinfo {author} {\bibfnamefont {A.}~\bibnamefont {Dr\'eau}},\ }\href
  {\doibase 10.1103/PhysRevLett.126.083602} {\bibfield  {journal} {\bibinfo
  {journal} {Phys. Rev. Lett.}\ }\textbf {\bibinfo {volume} {126}},\ \bibinfo
  {pages} {083602} (\bibinfo {year} {2021})}\BibitemShut {NoStop}%
\bibitem [{\citenamefont {Zhang}\ \emph {et~al.}(2020)\citenamefont {Zhang},
  \citenamefont {Cheng}, \citenamefont {Chou},\ and\ \citenamefont
  {Gali}}]{intro-Zhang}%
  \BibitemOpen
  \bibfield  {author} {\bibinfo {author} {\bibfnamefont {G.}~\bibnamefont
  {Zhang}}, \bibinfo {author} {\bibfnamefont {Y.}~\bibnamefont {Cheng}},
  \bibinfo {author} {\bibfnamefont {J.-P.}\ \bibnamefont {Chou}}, \ and\
  \bibinfo {author} {\bibfnamefont {A.}~\bibnamefont {Gali}},\ }\href {\doibase
  10.1063/5.0006075} {\bibfield  {journal} {\bibinfo  {journal} {Applied
  Physics Reviews}\ }\textbf {\bibinfo {volume} {7}},\ \bibinfo {pages}
  {031308} (\bibinfo {year} {2020})},\ \Eprint
  {http://arxiv.org/abs/https://doi.org/10.1063/5.0006075}
  {https://doi.org/10.1063/5.0006075} \BibitemShut {NoStop}%
\bibitem [{\citenamefont {Song}\ \emph {et~al.}(1990)\citenamefont {Song},
  \citenamefont {Zhan}, \citenamefont {Benson},\ and\ \citenamefont
  {Watkins}}]{GC1}%
  \BibitemOpen
  \bibfield  {author} {\bibinfo {author} {\bibfnamefont {L.~W.}\ \bibnamefont
  {Song}}, \bibinfo {author} {\bibfnamefont {X.~D.}\ \bibnamefont {Zhan}},
  \bibinfo {author} {\bibfnamefont {B.~W.}\ \bibnamefont {Benson}}, \ and\
  \bibinfo {author} {\bibfnamefont {G.~D.}\ \bibnamefont {Watkins}},\ }\href
  {\doibase 10.1103/PhysRevB.42.5765} {\bibfield  {journal} {\bibinfo
  {journal} {Phys. Rev. B}\ }\textbf {\bibinfo {volume} {42}},\ \bibinfo
  {pages} {5765} (\bibinfo {year} {1990})}\BibitemShut {NoStop}%
\bibitem [{\citenamefont {Docaj}\ and\ \citenamefont {Estreicher}(2012)}]{GC2}%
  \BibitemOpen
  \bibfield  {author} {\bibinfo {author} {\bibfnamefont {A.}~\bibnamefont
  {Docaj}}\ and\ \bibinfo {author} {\bibfnamefont {S.}~\bibnamefont
  {Estreicher}},\ }\href {\doibase https://doi.org/10.1016/j.physb.2011.08.029}
  {\bibfield  {journal} {\bibinfo  {journal} {Physica B: Condensed Matter}\
  }\textbf {\bibinfo {volume} {407}},\ \bibinfo {pages} {2981} (\bibinfo {year}
  {2012})},\ \bibinfo {note} {26th International Conference on Defects in
  Semiconductors}\BibitemShut {NoStop}%
\bibitem [{\citenamefont {Timerkaeva}\ \emph {et~al.}(2018)\citenamefont
  {Timerkaeva}, \citenamefont {Attaccalite}, \citenamefont {Brenet},
  \citenamefont {Caliste},\ and\ \citenamefont {Pochet}}]{GC3}%
  \BibitemOpen
  \bibfield  {author} {\bibinfo {author} {\bibfnamefont {D.}~\bibnamefont
  {Timerkaeva}}, \bibinfo {author} {\bibfnamefont {C.}~\bibnamefont
  {Attaccalite}}, \bibinfo {author} {\bibfnamefont {G.}~\bibnamefont {Brenet}},
  \bibinfo {author} {\bibfnamefont {D.}~\bibnamefont {Caliste}}, \ and\
  \bibinfo {author} {\bibfnamefont {P.}~\bibnamefont {Pochet}},\ }\href
  {\doibase 10.1063/1.5010269} {\bibfield  {journal} {\bibinfo  {journal}
  {Journal of Applied Physics}\ }\textbf {\bibinfo {volume} {123}},\ \bibinfo
  {pages} {161421} (\bibinfo {year} {2018})},\ \Eprint
  {http://arxiv.org/abs/1702.02334} {arXiv:1702.02334} \BibitemShut {NoStop}%
\bibitem [{\citenamefont {Wang}\ \emph {et~al.}(2014)\citenamefont {Wang},
  \citenamefont {Chroneos}, \citenamefont {Londos}, \citenamefont {Sgourou},\
  and\ \citenamefont {Schwingenschlögl}}]{Gcenter-HSE}%
  \BibitemOpen
  \bibfield  {author} {\bibinfo {author} {\bibfnamefont {H.}~\bibnamefont
  {Wang}}, \bibinfo {author} {\bibfnamefont {A.}~\bibnamefont {Chroneos}},
  \bibinfo {author} {\bibfnamefont {C.~A.}\ \bibnamefont {Londos}}, \bibinfo
  {author} {\bibfnamefont {E.~N.}\ \bibnamefont {Sgourou}}, \ and\ \bibinfo
  {author} {\bibfnamefont {U.}~\bibnamefont {Schwingenschlögl}},\ }\href
  {\doibase 10.1063/1.4875658} {\bibfield  {journal} {\bibinfo  {journal}
  {Journal of Applied Physics}\ }\textbf {\bibinfo {volume} {115}},\ \bibinfo
  {pages} {183509} (\bibinfo {year} {2014})}\BibitemShut {NoStop}%
\bibitem [{\citenamefont {Richie}\ \emph {et~al.}(2004)\citenamefont {Richie},
  \citenamefont {Kim}, \citenamefont {Barr}, \citenamefont {Hazzard},
  \citenamefont {Hennig},\ and\ \citenamefont {Wilkins}}]{WC-theory-Vdiscov}%
  \BibitemOpen
  \bibfield  {author} {\bibinfo {author} {\bibfnamefont {D.~A.}\ \bibnamefont
  {Richie}}, \bibinfo {author} {\bibfnamefont {J.}~\bibnamefont {Kim}},
  \bibinfo {author} {\bibfnamefont {S.~A.}\ \bibnamefont {Barr}}, \bibinfo
  {author} {\bibfnamefont {K.~R.~A.}\ \bibnamefont {Hazzard}}, \bibinfo
  {author} {\bibfnamefont {R.}~\bibnamefont {Hennig}}, \ and\ \bibinfo {author}
  {\bibfnamefont {J.~W.}\ \bibnamefont {Wilkins}},\ }\href {\doibase
  10.1103/PhysRevLett.92.045501} {\bibfield  {journal} {\bibinfo  {journal}
  {Phys. Rev. Lett.}\ }\textbf {\bibinfo {volume} {92}},\ \bibinfo {pages}
  {045501} (\bibinfo {year} {2004})}\BibitemShut {NoStop}%
\bibitem [{\citenamefont {Carvalho}\ \emph {et~al.}(2005)\citenamefont
  {Carvalho}, \citenamefont {Jones}, \citenamefont {Coutinho},\ and\
  \citenamefont {Briddon}}]{WC-theory-V1}%
  \BibitemOpen
  \bibfield  {author} {\bibinfo {author} {\bibfnamefont {A.}~\bibnamefont
  {Carvalho}}, \bibinfo {author} {\bibfnamefont {R.}~\bibnamefont {Jones}},
  \bibinfo {author} {\bibfnamefont {J.}~\bibnamefont {Coutinho}}, \ and\
  \bibinfo {author} {\bibfnamefont {P.~R.}\ \bibnamefont {Briddon}},\ }\href
  {\doibase 10.1103/PhysRevB.72.155208} {\bibfield  {journal} {\bibinfo
  {journal} {Phys. Rev. B}\ }\textbf {\bibinfo {volume} {72}},\ \bibinfo
  {pages} {155208} (\bibinfo {year} {2005})}\BibitemShut {NoStop}%
\bibitem [{\citenamefont {Santos}\ \emph {et~al.}(2016)\citenamefont {Santos},
  \citenamefont {Aboy}, \citenamefont {L{\'{o}}pez}, \citenamefont
  {Marqu{\'{e}}s},\ and\ \citenamefont {Pelaz}}]{WC-theory-V2}%
  \BibitemOpen
  \bibfield  {author} {\bibinfo {author} {\bibfnamefont {I.}~\bibnamefont
  {Santos}}, \bibinfo {author} {\bibfnamefont {M.}~\bibnamefont {Aboy}},
  \bibinfo {author} {\bibfnamefont {P.}~\bibnamefont {L{\'{o}}pez}}, \bibinfo
  {author} {\bibfnamefont {L.~A.}\ \bibnamefont {Marqu{\'{e}}s}}, \ and\
  \bibinfo {author} {\bibfnamefont {L.}~\bibnamefont {Pelaz}},\ }\href
  {\doibase 10.1088/0022-3727/49/7/075109} {\bibfield  {journal} {\bibinfo
  {journal} {Journal of Physics D: Applied Physics}\ }\textbf {\bibinfo
  {volume} {49}},\ \bibinfo {pages} {075109} (\bibinfo {year}
  {2016})}\BibitemShut {NoStop}%
\bibitem [{\citenamefont {Dhaliah}\ \emph {et~al.}(2022)\citenamefont
  {Dhaliah}, \citenamefont {Xiong}, \citenamefont {Sipahigil}, \citenamefont
  {Griffin},\ and\ \citenamefont {Hautier}}]{sinead-TC}%
  \BibitemOpen
  \bibfield  {author} {\bibinfo {author} {\bibfnamefont {D.}~\bibnamefont
  {Dhaliah}}, \bibinfo {author} {\bibfnamefont {Y.}~\bibnamefont {Xiong}},
  \bibinfo {author} {\bibfnamefont {A.}~\bibnamefont {Sipahigil}}, \bibinfo
  {author} {\bibfnamefont {S.~M.}\ \bibnamefont {Griffin}}, \ and\ \bibinfo
  {author} {\bibfnamefont {G.}~\bibnamefont {Hautier}},\ }\href@noop {}
  {\enquote {\bibinfo {title} {First principles study of the t-center in
  silicon},}\ } (\bibinfo {year} {2022}),\ \Eprint
  {http://arxiv.org/abs/2202.04149} {arXiv:2202.04149 [cond-mat.mtrl-sci]}
  \BibitemShut {NoStop}%
\bibitem [{\citenamefont {Safonov}\ \emph {et~al.}(1996)\citenamefont
  {Safonov}, \citenamefont {Lightowlers}, \citenamefont {Davies}, \citenamefont
  {Leary}, \citenamefont {Jones},\ and\ \citenamefont
  {\"Oberg}}]{tcenter-safonov}%
  \BibitemOpen
  \bibfield  {author} {\bibinfo {author} {\bibfnamefont {A.~N.}\ \bibnamefont
  {Safonov}}, \bibinfo {author} {\bibfnamefont {E.~C.}\ \bibnamefont
  {Lightowlers}}, \bibinfo {author} {\bibfnamefont {G.}~\bibnamefont {Davies}},
  \bibinfo {author} {\bibfnamefont {P.}~\bibnamefont {Leary}}, \bibinfo
  {author} {\bibfnamefont {R.}~\bibnamefont {Jones}}, \ and\ \bibinfo {author}
  {\bibfnamefont {S.}~\bibnamefont {\"Oberg}},\ }\href {\doibase
  10.1103/PhysRevLett.77.4812} {\bibfield  {journal} {\bibinfo  {journal}
  {Phys. Rev. Lett.}\ }\textbf {\bibinfo {volume} {77}},\ \bibinfo {pages}
  {4812} (\bibinfo {year} {1996})}\BibitemShut {NoStop}%
\bibitem [{sup()}]{supplement}%
  \BibitemOpen
  \href@noop {} {}\bibinfo {note} {The Supplemental Material at [URL will be
  inserted by publisher] contains further details about the defect structures
  and photoluminescence calculations.}\BibitemShut {Stop}%
\bibitem [{\citenamefont {Davies}(1989)}]{1989_GDavies_PR}%
  \BibitemOpen
  \bibfield  {author} {\bibinfo {author} {\bibfnamefont {G.}~\bibnamefont
  {Davies}},\ }\href {\doibase https://doi.org/10.1016/0370-1573(89)90064-1}
  {\bibfield  {journal} {\bibinfo  {journal} {Physics Reports}\ }\textbf
  {\bibinfo {volume} {176}},\ \bibinfo {pages} {83} (\bibinfo {year}
  {1989})}\BibitemShut {NoStop}%
\bibitem [{\citenamefont {Davies}\ \emph {et~al.}(1987)\citenamefont {Davies},
  \citenamefont {Lightowlers},\ and\ \citenamefont
  {Ciechanowska}}]{1987_GDavies_JPC}%
  \BibitemOpen
  \bibfield  {author} {\bibinfo {author} {\bibfnamefont {G.}~\bibnamefont
  {Davies}}, \bibinfo {author} {\bibfnamefont {E.~C.}\ \bibnamefont
  {Lightowlers}}, \ and\ \bibinfo {author} {\bibfnamefont {Z.~E.}\ \bibnamefont
  {Ciechanowska}},\ }\href {\doibase 10.1088/0022-3719/20/2/003} {\bibfield
  {journal} {\bibinfo  {journal} {Journal of Physics C: Solid State Physics}\
  }\textbf {\bibinfo {volume} {20}},\ \bibinfo {pages} {191} (\bibinfo {year}
  {1987})}\BibitemShut {NoStop}%
\bibitem [{\citenamefont {Tkachev}\ and\ \citenamefont
  {Mudryi}(1978)}]{1978_Tkachev_JAS}%
  \BibitemOpen
  \bibfield  {author} {\bibinfo {author} {\bibfnamefont {V.~D.}\ \bibnamefont
  {Tkachev}}\ and\ \bibinfo {author} {\bibfnamefont {A.~V.}\ \bibnamefont
  {Mudryi}},\ }\href {\doibase 10.1007/BF00613548} {\bibfield  {journal}
  {\bibinfo  {journal} {Journal of Applied Spectroscopy}\ }\textbf {\bibinfo
  {volume} {29}},\ \bibinfo {pages} {1485} (\bibinfo {year}
  {1978})}\BibitemShut {NoStop}%
\bibitem [{\citenamefont {B\"urger}\ \emph {et~al.}(1984)\citenamefont
  {B\"urger}, \citenamefont {Thonke}, \citenamefont {Sauer},\ and\
  \citenamefont {Pensl}}]{1984_Burger_PRL}%
  \BibitemOpen
  \bibfield  {author} {\bibinfo {author} {\bibfnamefont {N.}~\bibnamefont
  {B\"urger}}, \bibinfo {author} {\bibfnamefont {K.}~\bibnamefont {Thonke}},
  \bibinfo {author} {\bibfnamefont {R.}~\bibnamefont {Sauer}}, \ and\ \bibinfo
  {author} {\bibfnamefont {G.}~\bibnamefont {Pensl}},\ }\href {\doibase
  10.1103/PhysRevLett.52.1645} {\bibfield  {journal} {\bibinfo  {journal}
  {Phys. Rev. Lett.}\ }\textbf {\bibinfo {volume} {52}},\ \bibinfo {pages}
  {1645} (\bibinfo {year} {1984})}\BibitemShut {NoStop}%
\bibitem [{\citenamefont {Coomer}\ \emph {et~al.}(1999)\citenamefont {Coomer},
  \citenamefont {Goss}, \citenamefont {Jones}, \citenamefont {Öberg},\ and\
  \citenamefont {Briddon}}]{WC-theory1}%
  \BibitemOpen
  \bibfield  {author} {\bibinfo {author} {\bibfnamefont {B.}~\bibnamefont
  {Coomer}}, \bibinfo {author} {\bibfnamefont {J.}~\bibnamefont {Goss}},
  \bibinfo {author} {\bibfnamefont {R.}~\bibnamefont {Jones}}, \bibinfo
  {author} {\bibfnamefont {S.}~\bibnamefont {Öberg}}, \ and\ \bibinfo {author}
  {\bibfnamefont {P.}~\bibnamefont {Briddon}},\ }\href {\doibase
  https://doi.org/10.1016/S0921-4526(99)00538-4} {\bibfield  {journal}
  {\bibinfo  {journal} {Physica B: Condensed Matter}\ }\textbf {\bibinfo
  {volume} {273-274}},\ \bibinfo {pages} {505} (\bibinfo {year}
  {1999})}\BibitemShut {NoStop}%
\bibitem [{\citenamefont {Gharaibeh}\ \emph {et~al.}(1999)\citenamefont
  {Gharaibeh}, \citenamefont {Estreicher},\ and\ \citenamefont
  {Fedders}}]{WC-theory2}%
  \BibitemOpen
  \bibfield  {author} {\bibinfo {author} {\bibfnamefont {M.}~\bibnamefont
  {Gharaibeh}}, \bibinfo {author} {\bibfnamefont {S.}~\bibnamefont
  {Estreicher}}, \ and\ \bibinfo {author} {\bibfnamefont {P.}~\bibnamefont
  {Fedders}},\ }\href {\doibase https://doi.org/10.1016/S0921-4526(99)00566-9}
  {\bibfield  {journal} {\bibinfo  {journal} {Physica B: Condensed Matter}\
  }\textbf {\bibinfo {volume} {273-274}},\ \bibinfo {pages} {532} (\bibinfo
  {year} {1999})}\BibitemShut {NoStop}%
\bibitem [{\citenamefont {Jones}\ \emph {et~al.}(2002)\citenamefont {Jones},
  \citenamefont {Eberlein}, \citenamefont {Pinho}, \citenamefont {Coomer},
  \citenamefont {Goss}, \citenamefont {Briddon},\ and\ \citenamefont
  {Öberg}}]{WC-theory3}%
  \BibitemOpen
  \bibfield  {author} {\bibinfo {author} {\bibfnamefont {R.}~\bibnamefont
  {Jones}}, \bibinfo {author} {\bibfnamefont {T.}~\bibnamefont {Eberlein}},
  \bibinfo {author} {\bibfnamefont {N.}~\bibnamefont {Pinho}}, \bibinfo
  {author} {\bibfnamefont {B.}~\bibnamefont {Coomer}}, \bibinfo {author}
  {\bibfnamefont {J.}~\bibnamefont {Goss}}, \bibinfo {author} {\bibfnamefont
  {P.}~\bibnamefont {Briddon}}, \ and\ \bibinfo {author} {\bibfnamefont
  {S.}~\bibnamefont {Öberg}},\ }\href {\doibase
  https://doi.org/10.1016/S0168-583X(01)00872-2} {\bibfield  {journal}
  {\bibinfo  {journal} {Nuclear Instruments and Methods in Physics Research
  Section B: Beam Interactions with Materials and Atoms}\ }\textbf {\bibinfo
  {volume} {186}},\ \bibinfo {pages} {10} (\bibinfo {year} {2002})}\BibitemShut
  {NoStop}%
\bibitem [{\citenamefont {Rasband}\ \emph {et~al.}(1996)\citenamefont
  {Rasband}, \citenamefont {Clancy},\ and\ \citenamefont
  {Thompson}}]{WC-theory4}%
  \BibitemOpen
  \bibfield  {author} {\bibinfo {author} {\bibfnamefont {P.~B.}\ \bibnamefont
  {Rasband}}, \bibinfo {author} {\bibfnamefont {P.}~\bibnamefont {Clancy}}, \
  and\ \bibinfo {author} {\bibfnamefont {M.~O.}\ \bibnamefont {Thompson}},\
  }\href {\doibase 10.1063/1.362632} {\bibfield  {journal} {\bibinfo  {journal}
  {Journal of Applied Physics}\ }\textbf {\bibinfo {volume} {79}},\ \bibinfo
  {pages} {8998} (\bibinfo {year} {1996})}\BibitemShut {NoStop}%
\bibitem [{\citenamefont {Lopez}\ and\ \citenamefont
  {Fiorentini}(2004)}]{WC-theory-II}%
  \BibitemOpen
  \bibfield  {author} {\bibinfo {author} {\bibfnamefont {G.~M.}\ \bibnamefont
  {Lopez}}\ and\ \bibinfo {author} {\bibfnamefont {V.}~\bibnamefont
  {Fiorentini}},\ }\href {\doibase 10.1103/PhysRevB.69.155206} {\bibfield
  {journal} {\bibinfo  {journal} {Phys. Rev. B}\ }\textbf {\bibinfo {volume}
  {69}},\ \bibinfo {pages} {155206} (\bibinfo {year} {2004})}\BibitemShut
  {NoStop}%
\bibitem [{\citenamefont {Baron}\ \emph {et~al.}(2021)\citenamefont {Baron},
  \citenamefont {Durand}, \citenamefont {Udvarhelyi}, \citenamefont {Herzig},
  \citenamefont {Khoury}, \citenamefont {Pezzagna}, \citenamefont {Meijer},
  \citenamefont {Robert-Philip}, \citenamefont {Abbarchi}, \citenamefont
  {Hartmann}, \citenamefont {Mazzocchi}, \citenamefont {Gérard}, \citenamefont
  {Gali}, \citenamefont {Jacques}, \citenamefont {Cassabois},\ and\
  \citenamefont {Dréau}}]{WCsingle}%
  \BibitemOpen
  \bibfield  {author} {\bibinfo {author} {\bibfnamefont {Y.}~\bibnamefont
  {Baron}}, \bibinfo {author} {\bibfnamefont {A.}~\bibnamefont {Durand}},
  \bibinfo {author} {\bibfnamefont {P.}~\bibnamefont {Udvarhelyi}}, \bibinfo
  {author} {\bibfnamefont {T.}~\bibnamefont {Herzig}}, \bibinfo {author}
  {\bibfnamefont {M.}~\bibnamefont {Khoury}}, \bibinfo {author} {\bibfnamefont
  {S.}~\bibnamefont {Pezzagna}}, \bibinfo {author} {\bibfnamefont
  {J.}~\bibnamefont {Meijer}}, \bibinfo {author} {\bibfnamefont
  {I.}~\bibnamefont {Robert-Philip}}, \bibinfo {author} {\bibfnamefont
  {M.}~\bibnamefont {Abbarchi}}, \bibinfo {author} {\bibfnamefont {J.-M.}\
  \bibnamefont {Hartmann}}, \bibinfo {author} {\bibfnamefont {V.}~\bibnamefont
  {Mazzocchi}}, \bibinfo {author} {\bibfnamefont {J.-M.}\ \bibnamefont
  {Gérard}}, \bibinfo {author} {\bibfnamefont {A.}~\bibnamefont {Gali}},
  \bibinfo {author} {\bibfnamefont {V.}~\bibnamefont {Jacques}}, \bibinfo
  {author} {\bibfnamefont {G.}~\bibnamefont {Cassabois}}, \ and\ \bibinfo
  {author} {\bibfnamefont {A.}~\bibnamefont {Dréau}},\ }\href@noop {}
  {\enquote {\bibinfo {title} {Detection of single w-centers in silicon},}\ }
  (\bibinfo {year} {2021}),\ \Eprint {http://arxiv.org/abs/arXiv:2108.04283}
  {arXiv:2108.04283} \BibitemShut {NoStop}%
\bibitem [{\citenamefont {Bergeron}\ \emph
  {et~al.}(2020{\natexlab{b}})\citenamefont {Bergeron}, \citenamefont
  {Chartrand}, \citenamefont {Kurkjian}, \citenamefont {Morse}, \citenamefont
  {Riemann}, \citenamefont {Abrosimov}, \citenamefont {Becker}, \citenamefont
  {Pohl}, \citenamefont {Thewalt},\ and\ \citenamefont
  {Simmons}}]{tcenter-28Si}%
  \BibitemOpen
  \bibfield  {author} {\bibinfo {author} {\bibfnamefont {L.}~\bibnamefont
  {Bergeron}}, \bibinfo {author} {\bibfnamefont {C.}~\bibnamefont {Chartrand}},
  \bibinfo {author} {\bibfnamefont {A.~T.~K.}\ \bibnamefont {Kurkjian}},
  \bibinfo {author} {\bibfnamefont {K.~J.}\ \bibnamefont {Morse}}, \bibinfo
  {author} {\bibfnamefont {H.}~\bibnamefont {Riemann}}, \bibinfo {author}
  {\bibfnamefont {N.~V.}\ \bibnamefont {Abrosimov}}, \bibinfo {author}
  {\bibfnamefont {P.}~\bibnamefont {Becker}}, \bibinfo {author} {\bibfnamefont
  {H.~J.}\ \bibnamefont {Pohl}}, \bibinfo {author} {\bibfnamefont {M.~L.~W.}\
  \bibnamefont {Thewalt}}, \ and\ \bibinfo {author} {\bibfnamefont
  {S.}~\bibnamefont {Simmons}},\ }\href@noop {} {\enquote {\bibinfo {title}
  {Characterization of the t center in $^{28}$si},}\ } (\bibinfo {year}
  {2020}{\natexlab{b}}),\ \Eprint {http://arxiv.org/abs/2006.08794}
  {arXiv:2006.08794 [cond-mat.mtrl-sci]} \BibitemShut {NoStop}%
\bibitem [{\citenamefont {Irion}\ \emph {et~al.}(1985)\citenamefont {Irion},
  \citenamefont {Burger}, \citenamefont {Thonke},\ and\ \citenamefont
  {Sauer}}]{tcenter-irion}%
  \BibitemOpen
  \bibfield  {author} {\bibinfo {author} {\bibfnamefont {E.}~\bibnamefont
  {Irion}}, \bibinfo {author} {\bibfnamefont {N.}~\bibnamefont {Burger}},
  \bibinfo {author} {\bibfnamefont {K.}~\bibnamefont {Thonke}}, \ and\ \bibinfo
  {author} {\bibfnamefont {R.}~\bibnamefont {Sauer}},\ }\href {\doibase
  10.1088/0022-3719/18/26/018} {\bibfield  {journal} {\bibinfo  {journal}
  {Journal of Physics C: Solid State Physics}\ }\textbf {\bibinfo {volume}
  {18}},\ \bibinfo {pages} {5069} (\bibinfo {year} {1985})}\BibitemShut
  {NoStop}%
\bibitem [{\citenamefont {Beaufils}\ \emph {et~al.}(2018)\citenamefont
  {Beaufils}, \citenamefont {Redjem}, \citenamefont {Rousseau}, \citenamefont
  {Jacques}, \citenamefont {Kuznetsov}, \citenamefont {Raynaud}, \citenamefont
  {Voisin}, \citenamefont {Benali}, \citenamefont {Herzig}, \citenamefont
  {Pezzagna}, \citenamefont {Meijer}, \citenamefont {Abbarchi},\ and\
  \citenamefont {Cassabois}}]{GCensemble}%
  \BibitemOpen
  \bibfield  {author} {\bibinfo {author} {\bibfnamefont {C.}~\bibnamefont
  {Beaufils}}, \bibinfo {author} {\bibfnamefont {W.}~\bibnamefont {Redjem}},
  \bibinfo {author} {\bibfnamefont {E.}~\bibnamefont {Rousseau}}, \bibinfo
  {author} {\bibfnamefont {V.}~\bibnamefont {Jacques}}, \bibinfo {author}
  {\bibfnamefont {A.~Y.}\ \bibnamefont {Kuznetsov}}, \bibinfo {author}
  {\bibfnamefont {C.}~\bibnamefont {Raynaud}}, \bibinfo {author} {\bibfnamefont
  {C.}~\bibnamefont {Voisin}}, \bibinfo {author} {\bibfnamefont
  {A.}~\bibnamefont {Benali}}, \bibinfo {author} {\bibfnamefont
  {T.}~\bibnamefont {Herzig}}, \bibinfo {author} {\bibfnamefont
  {S.}~\bibnamefont {Pezzagna}}, \bibinfo {author} {\bibfnamefont
  {J.}~\bibnamefont {Meijer}}, \bibinfo {author} {\bibfnamefont
  {M.}~\bibnamefont {Abbarchi}}, \ and\ \bibinfo {author} {\bibfnamefont
  {G.}~\bibnamefont {Cassabois}},\ }\href {\doibase 10.1103/PhysRevB.97.035303}
  {\bibfield  {journal} {\bibinfo  {journal} {Phys. Rev. B}\ }\textbf {\bibinfo
  {volume} {97}},\ \bibinfo {pages} {035303} (\bibinfo {year}
  {2018})}\BibitemShut {NoStop}%
\bibitem [{\citenamefont {Schenkel}\ \emph {et~al.}(2022)\citenamefont
  {Schenkel}, \citenamefont {Redjem}, \citenamefont {Persaud}, \citenamefont
  {Liu}, \citenamefont {Seidl}, \citenamefont {Amsellem}, \citenamefont
  {Kanté},\ and\ \citenamefont {Ji}}]{qubit-exploration}%
  \BibitemOpen
  \bibfield  {author} {\bibinfo {author} {\bibfnamefont {T.}~\bibnamefont
  {Schenkel}}, \bibinfo {author} {\bibfnamefont {W.}~\bibnamefont {Redjem}},
  \bibinfo {author} {\bibfnamefont {A.}~\bibnamefont {Persaud}}, \bibinfo
  {author} {\bibfnamefont {W.}~\bibnamefont {Liu}}, \bibinfo {author}
  {\bibfnamefont {P.~A.}\ \bibnamefont {Seidl}}, \bibinfo {author}
  {\bibfnamefont {A.~J.}\ \bibnamefont {Amsellem}}, \bibinfo {author}
  {\bibfnamefont {B.}~\bibnamefont {Kanté}}, \ and\ \bibinfo {author}
  {\bibfnamefont {Q.}~\bibnamefont {Ji}},\ }\href {\doibase
  10.3390/qubs6010013} {\bibfield  {journal} {\bibinfo  {journal} {Quantum Beam
  Science}\ }\textbf {\bibinfo {volume} {6}} (\bibinfo {year} {2022}),\
  10.3390/qubs6010013}\BibitemShut {NoStop}%
\bibitem [{\citenamefont {O'Donnell}\ \emph {et~al.}(1983)\citenamefont
  {O'Donnell}, \citenamefont {Lee},\ and\ \citenamefont {Watkins}}]{zfs-exp}%
  \BibitemOpen
  \bibfield  {author} {\bibinfo {author} {\bibfnamefont {K.}~\bibnamefont
  {O'Donnell}}, \bibinfo {author} {\bibfnamefont {K.}~\bibnamefont {Lee}}, \
  and\ \bibinfo {author} {\bibfnamefont {G.}~\bibnamefont {Watkins}},\ }\href
  {\doibase https://doi.org/10.1016/0378-4363(83)90256-5} {\bibfield  {journal}
  {\bibinfo  {journal} {Physica B+C}\ }\textbf {\bibinfo {volume} {116}},\
  \bibinfo {pages} {258} (\bibinfo {year} {1983})}\BibitemShut {NoStop}%
\bibitem [{\citenamefont {Leary}\ \emph {et~al.}(1998)\citenamefont {Leary},
  \citenamefont {Jones},\ and\ \citenamefont {\"Oberg}}]{ch-defects}%
  \BibitemOpen
  \bibfield  {author} {\bibinfo {author} {\bibfnamefont {P.}~\bibnamefont
  {Leary}}, \bibinfo {author} {\bibfnamefont {R.}~\bibnamefont {Jones}}, \ and\
  \bibinfo {author} {\bibfnamefont {S.}~\bibnamefont {\"Oberg}},\ }\href
  {\doibase 10.1103/PhysRevB.57.3887} {\bibfield  {journal} {\bibinfo
  {journal} {Phys. Rev. B}\ }\textbf {\bibinfo {volume} {57}},\ \bibinfo
  {pages} {3887} (\bibinfo {year} {1998})}\BibitemShut {NoStop}%
\bibitem [{\citenamefont {Kresse}\ and\ \citenamefont {Hafner}(1993)}]{vasp1}%
  \BibitemOpen
  \bibfield  {author} {\bibinfo {author} {\bibfnamefont {G.}~\bibnamefont
  {Kresse}}\ and\ \bibinfo {author} {\bibfnamefont {J.}~\bibnamefont
  {Hafner}},\ }\href {\doibase 10.1103/PhysRevB.47.558} {\bibfield  {journal}
  {\bibinfo  {journal} {Phys. Rev. B}\ }\textbf {\bibinfo {volume} {47}},\
  \bibinfo {pages} {558} (\bibinfo {year} {1993})}\BibitemShut {NoStop}%
\bibitem [{\citenamefont {Kresse}\ and\ \citenamefont {Hafner}(1994)}]{vasp2}%
  \BibitemOpen
  \bibfield  {author} {\bibinfo {author} {\bibfnamefont {G.}~\bibnamefont
  {Kresse}}\ and\ \bibinfo {author} {\bibfnamefont {J.}~\bibnamefont
  {Hafner}},\ }\href {\doibase 10.1103/PhysRevB.49.14251} {\bibfield  {journal}
  {\bibinfo  {journal} {Phys. Rev. B}\ }\textbf {\bibinfo {volume} {49}},\
  \bibinfo {pages} {14251} (\bibinfo {year} {1994})}\BibitemShut {NoStop}%
\bibitem [{\citenamefont {Kresse}\ and\ \citenamefont
  {Furthmüller}(1996)}]{vasp3}%
  \BibitemOpen
  \bibfield  {author} {\bibinfo {author} {\bibfnamefont {G.}~\bibnamefont
  {Kresse}}\ and\ \bibinfo {author} {\bibfnamefont {J.}~\bibnamefont
  {Furthmüller}},\ }\href {\doibase
  https://doi.org/10.1016/0927-0256(96)00008-0} {\bibfield  {journal} {\bibinfo
   {journal} {Computational Materials Science}\ }\textbf {\bibinfo {volume}
  {6}},\ \bibinfo {pages} {15} (\bibinfo {year} {1996})}\BibitemShut {NoStop}%
\bibitem [{\citenamefont {Kresse}\ and\ \citenamefont
  {Furthm\"uller}(1996)}]{vasp4}%
  \BibitemOpen
  \bibfield  {author} {\bibinfo {author} {\bibfnamefont {G.}~\bibnamefont
  {Kresse}}\ and\ \bibinfo {author} {\bibfnamefont {J.}~\bibnamefont
  {Furthm\"uller}},\ }\href {\doibase 10.1103/PhysRevB.54.11169} {\bibfield
  {journal} {\bibinfo  {journal} {Phys. Rev. B}\ }\textbf {\bibinfo {volume}
  {54}},\ \bibinfo {pages} {11169} (\bibinfo {year} {1996})}\BibitemShut
  {NoStop}%
\bibitem [{\citenamefont {Krukau}\ \emph {et~al.}(2006)\citenamefont {Krukau},
  \citenamefont {Vydrov}, \citenamefont {Izmaylov},\ and\ \citenamefont
  {Scuseria}}]{hse}%
  \BibitemOpen
  \bibfield  {author} {\bibinfo {author} {\bibfnamefont {A.~V.}\ \bibnamefont
  {Krukau}}, \bibinfo {author} {\bibfnamefont {O.~A.}\ \bibnamefont {Vydrov}},
  \bibinfo {author} {\bibfnamefont {A.~F.}\ \bibnamefont {Izmaylov}}, \ and\
  \bibinfo {author} {\bibfnamefont {G.~E.}\ \bibnamefont {Scuseria}},\ }\href
  {\doibase 10.1063/1.2404663} {\bibfield  {journal} {\bibinfo  {journal} {The
  Journal of Chemical Physics}\ }\textbf {\bibinfo {volume} {125}},\ \bibinfo
  {pages} {224106} (\bibinfo {year} {2006})}\BibitemShut {NoStop}%
\bibitem [{\citenamefont {Gali}\ \emph {et~al.}(2009)\citenamefont {Gali},
  \citenamefont {Janz\'en}, \citenamefont {De\'ak}, \citenamefont {Kresse},\
  and\ \citenamefont {Kaxiras}}]{gali-constrainedNV}%
  \BibitemOpen
  \bibfield  {author} {\bibinfo {author} {\bibfnamefont {A.}~\bibnamefont
  {Gali}}, \bibinfo {author} {\bibfnamefont {E.}~\bibnamefont {Janz\'en}},
  \bibinfo {author} {\bibfnamefont {P.}~\bibnamefont {De\'ak}}, \bibinfo
  {author} {\bibfnamefont {G.}~\bibnamefont {Kresse}}, \ and\ \bibinfo {author}
  {\bibfnamefont {E.}~\bibnamefont {Kaxiras}},\ }\href {\doibase
  10.1103/PhysRevLett.103.186404} {\bibfield  {journal} {\bibinfo  {journal}
  {Phys. Rev. Lett.}\ }\textbf {\bibinfo {volume} {103}},\ \bibinfo {pages}
  {186404} (\bibinfo {year} {2009})}\BibitemShut {NoStop}%
\bibitem [{\citenamefont {Alkauskas}\ \emph {et~al.}(2014)\citenamefont
  {Alkauskas}, \citenamefont {Buckley}, \citenamefont {Awschalom},\ and\
  \citenamefont {de~Walle}}]{Alkauskas_2014}%
  \BibitemOpen
  \bibfield  {author} {\bibinfo {author} {\bibfnamefont {A.}~\bibnamefont
  {Alkauskas}}, \bibinfo {author} {\bibfnamefont {B.~B.}\ \bibnamefont
  {Buckley}}, \bibinfo {author} {\bibfnamefont {D.~D.}\ \bibnamefont
  {Awschalom}}, \ and\ \bibinfo {author} {\bibfnamefont {C.~G.~V.}\
  \bibnamefont {de~Walle}},\ }\href {\doibase 10.1088/1367-2630/16/7/073026}
  {\bibfield  {journal} {\bibinfo  {journal} {New Journal of Physics}\ }\textbf
  {\bibinfo {volume} {16}},\ \bibinfo {pages} {073026} (\bibinfo {year}
  {2014})}\BibitemShut {NoStop}%
\bibitem [{\citenamefont {Thonke}\ \emph {et~al.}(1981)\citenamefont {Thonke},
  \citenamefont {Klemisch}, \citenamefont {Weber},\ and\ \citenamefont
  {Sauer}}]{gcenter-eline}%
  \BibitemOpen
  \bibfield  {author} {\bibinfo {author} {\bibfnamefont {K.}~\bibnamefont
  {Thonke}}, \bibinfo {author} {\bibfnamefont {H.}~\bibnamefont {Klemisch}},
  \bibinfo {author} {\bibfnamefont {J.}~\bibnamefont {Weber}}, \ and\ \bibinfo
  {author} {\bibfnamefont {R.}~\bibnamefont {Sauer}},\ }\href {\doibase
  10.1103/PhysRevB.24.5874} {\bibfield  {journal} {\bibinfo  {journal} {Phys.
  Rev. B}\ }\textbf {\bibinfo {volume} {24}},\ \bibinfo {pages} {5874}
  (\bibinfo {year} {1981})}\BibitemShut {NoStop}%
\bibitem [{\citenamefont {MacQuarrie}\ \emph {et~al.}(2021)\citenamefont
  {MacQuarrie}, \citenamefont {Chartrand}, \citenamefont {Higginbottom},
  \citenamefont {Morse}, \citenamefont {Karasyuk}, \citenamefont {Roorda},\
  and\ \citenamefont {Simmons}}]{tcenter-lvm}%
  \BibitemOpen
  \bibfield  {author} {\bibinfo {author} {\bibfnamefont {E.~R.}\ \bibnamefont
  {MacQuarrie}}, \bibinfo {author} {\bibfnamefont {C.}~\bibnamefont
  {Chartrand}}, \bibinfo {author} {\bibfnamefont {D.~B.}\ \bibnamefont
  {Higginbottom}}, \bibinfo {author} {\bibfnamefont {K.~J.}\ \bibnamefont
  {Morse}}, \bibinfo {author} {\bibfnamefont {V.~A.}\ \bibnamefont {Karasyuk}},
  \bibinfo {author} {\bibfnamefont {S.}~\bibnamefont {Roorda}}, \ and\ \bibinfo
  {author} {\bibfnamefont {S.}~\bibnamefont {Simmons}},\ }\href {\doibase
  10.48550/ARXIV.2103.03998} {\enquote {\bibinfo {title} {T centres in photonic
  silicon-on-insulator material},}\ } (\bibinfo {year} {2021})\BibitemShut
  {NoStop}%
\bibitem [{\citenamefont {Huang}\ \emph {et~al.}(2021)\citenamefont {Huang},
  \citenamefont {Sarihan}, \citenamefont {Kang}, \citenamefont {Liang},
  \citenamefont {Liu},\ and\ \citenamefont {Wong}}]{tcenter-SOI}%
  \BibitemOpen
  \bibfield  {author} {\bibinfo {author} {\bibfnamefont {J.}~\bibnamefont
  {Huang}}, \bibinfo {author} {\bibfnamefont {M.~C.}\ \bibnamefont {Sarihan}},
  \bibinfo {author} {\bibfnamefont {J.~H.}\ \bibnamefont {Kang}}, \bibinfo
  {author} {\bibfnamefont {B.}~\bibnamefont {Liang}}, \bibinfo {author}
  {\bibfnamefont {W.}~\bibnamefont {Liu}}, \ and\ \bibinfo {author}
  {\bibfnamefont {C.~W.}\ \bibnamefont {Wong}},\ }in\ \href
  {http://opg.optica.org/abstract.cfm?URI=LS-2021-JTu7A.2} {\emph {\bibinfo
  {booktitle} {Frontiers in Optics $+$ Laser Science 2021}}}\ (\bibinfo
  {publisher} {Optica Publishing Group},\ \bibinfo {year} {2021})\ p.\ \bibinfo
  {pages} {JTu7A.2}\BibitemShut {NoStop}%
\bibitem [{\citenamefont {Prabhu}\ \emph {et~al.}(2022)\citenamefont {Prabhu},
  \citenamefont {Errando-Herranz}, \citenamefont {De~Santis}, \citenamefont
  {Christen}, \citenamefont {Chen},\ and\ \citenamefont
  {Englund}}]{gcenter-mit-waveguide}%
  \BibitemOpen
  \bibfield  {author} {\bibinfo {author} {\bibfnamefont {M.}~\bibnamefont
  {Prabhu}}, \bibinfo {author} {\bibfnamefont {C.}~\bibnamefont
  {Errando-Herranz}}, \bibinfo {author} {\bibfnamefont {L.}~\bibnamefont
  {De~Santis}}, \bibinfo {author} {\bibfnamefont {I.}~\bibnamefont {Christen}},
  \bibinfo {author} {\bibfnamefont {C.}~\bibnamefont {Chen}}, \ and\ \bibinfo
  {author} {\bibfnamefont {D.~R.}\ \bibnamefont {Englund}},\ }\href {\doibase
  10.48550/ARXIV.2202.02342} {\enquote {\bibinfo {title} {Individually
  addressable artificial atoms in silicon photonics},}\ } (\bibinfo {year}
  {2022})\BibitemShut {NoStop}%
\bibitem [{\citenamefont {Yan}\ \emph {et~al.}(2021)\citenamefont {Yan},
  \citenamefont {Gitt}, \citenamefont {Lin}, \citenamefont {Witt},
  \citenamefont {Abdolahi}, \citenamefont {Afifi}, \citenamefont {Azem},
  \citenamefont {Darcie}, \citenamefont {Wu}, \citenamefont {Awan},
  \citenamefont {Mitchell}, \citenamefont {Pfenning}, \citenamefont
  {Chrostowski},\ and\ \citenamefont {Young}}]{tcenter-quantum-computing}%
  \BibitemOpen
  \bibfield  {author} {\bibinfo {author} {\bibfnamefont {X.}~\bibnamefont
  {Yan}}, \bibinfo {author} {\bibfnamefont {S.}~\bibnamefont {Gitt}}, \bibinfo
  {author} {\bibfnamefont {B.}~\bibnamefont {Lin}}, \bibinfo {author}
  {\bibfnamefont {D.}~\bibnamefont {Witt}}, \bibinfo {author} {\bibfnamefont
  {M.}~\bibnamefont {Abdolahi}}, \bibinfo {author} {\bibfnamefont
  {A.}~\bibnamefont {Afifi}}, \bibinfo {author} {\bibfnamefont
  {A.}~\bibnamefont {Azem}}, \bibinfo {author} {\bibfnamefont {A.}~\bibnamefont
  {Darcie}}, \bibinfo {author} {\bibfnamefont {J.}~\bibnamefont {Wu}}, \bibinfo
  {author} {\bibfnamefont {K.}~\bibnamefont {Awan}}, \bibinfo {author}
  {\bibfnamefont {M.}~\bibnamefont {Mitchell}}, \bibinfo {author}
  {\bibfnamefont {A.}~\bibnamefont {Pfenning}}, \bibinfo {author}
  {\bibfnamefont {L.}~\bibnamefont {Chrostowski}}, \ and\ \bibinfo {author}
  {\bibfnamefont {J.~F.}\ \bibnamefont {Young}},\ }\href {\doibase
  10.1063/5.0049372} {\bibfield  {journal} {\bibinfo  {journal} {APL
  Photonics}\ }\textbf {\bibinfo {volume} {6}},\ \bibinfo {pages} {070901}
  (\bibinfo {year} {2021})},\ \Eprint
  {http://arxiv.org/abs/https://doi.org/10.1063/5.0049372}
  {https://doi.org/10.1063/5.0049372} \BibitemShut {NoStop}%
\bibitem [{\citenamefont {Pezzagna}\ and\ \citenamefont
  {Meijer}(2021)}]{qubit-diamond}%
  \BibitemOpen
  \bibfield  {author} {\bibinfo {author} {\bibfnamefont {S.}~\bibnamefont
  {Pezzagna}}\ and\ \bibinfo {author} {\bibfnamefont {J.}~\bibnamefont
  {Meijer}},\ }\href {\doibase 10.1063/5.0007444} {\bibfield  {journal}
  {\bibinfo  {journal} {Applied Physics Reviews}\ }\textbf {\bibinfo {volume}
  {8}},\ \bibinfo {pages} {011308} (\bibinfo {year} {2021})},\ \Eprint
  {http://arxiv.org/abs/https://doi.org/10.1063/5.0007444}
  {https://doi.org/10.1063/5.0007444} \BibitemShut {NoStop}%
\bibitem [{\citenamefont {Arroyo-Camejo}\ \emph {et~al.}(2014)\citenamefont
  {Arroyo-Camejo}, \citenamefont {Lazariev}, \citenamefont {Hell},\ and\
  \citenamefont {Balasubramanian}}]{qubit1}%
  \BibitemOpen
  \bibfield  {author} {\bibinfo {author} {\bibfnamefont {S.}~\bibnamefont
  {Arroyo-Camejo}}, \bibinfo {author} {\bibfnamefont {A.}~\bibnamefont
  {Lazariev}}, \bibinfo {author} {\bibfnamefont {S.~W.}\ \bibnamefont {Hell}},
  \ and\ \bibinfo {author} {\bibfnamefont {G.}~\bibnamefont
  {Balasubramanian}},\ }\href {\doibase 10.1038/ncomms5870} {\bibfield
  {journal} {\bibinfo  {journal} {Nature Communications}\ }\textbf {\bibinfo
  {volume} {5}},\ \bibinfo {pages} {4870} (\bibinfo {year} {2014})}\BibitemShut
  {NoStop}%
\bibitem [{\citenamefont {Zu}\ \emph {et~al.}(2014)\citenamefont {Zu},
  \citenamefont {Wang}, \citenamefont {He}, \citenamefont {Zhang},
  \citenamefont {Dai}, \citenamefont {Wang},\ and\ \citenamefont
  {Duan}}]{qubit2}%
  \BibitemOpen
  \bibfield  {author} {\bibinfo {author} {\bibfnamefont {C.}~\bibnamefont
  {Zu}}, \bibinfo {author} {\bibfnamefont {W.-B.}\ \bibnamefont {Wang}},
  \bibinfo {author} {\bibfnamefont {L.}~\bibnamefont {He}}, \bibinfo {author}
  {\bibfnamefont {W.-G.}\ \bibnamefont {Zhang}}, \bibinfo {author}
  {\bibfnamefont {C.-Y.}\ \bibnamefont {Dai}}, \bibinfo {author} {\bibfnamefont
  {F.}~\bibnamefont {Wang}}, \ and\ \bibinfo {author} {\bibfnamefont {L.-M.}\
  \bibnamefont {Duan}},\ }\href {\doibase 10.1038/nature13729} {\bibfield
  {journal} {\bibinfo  {journal} {Nature}\ }\textbf {\bibinfo {volume} {514}},\
  \bibinfo {pages} {72} (\bibinfo {year} {2014})}\BibitemShut {NoStop}%
\bibitem [{\citenamefont {Wu}\ \emph {et~al.}(2014)\citenamefont {Wu},
  \citenamefont {Ward}, \citenamefont {Prance}, \citenamefont {Kim},
  \citenamefont {Gamble}, \citenamefont {Mohr}, \citenamefont {Shi},
  \citenamefont {Savage}, \citenamefont {Lagally}, \citenamefont {Friesen},
  \citenamefont {Coppersmith},\ and\ \citenamefont {Eriksson}}]{qubit3}%
  \BibitemOpen
  \bibfield  {author} {\bibinfo {author} {\bibfnamefont {X.}~\bibnamefont
  {Wu}}, \bibinfo {author} {\bibfnamefont {D.~R.}\ \bibnamefont {Ward}},
  \bibinfo {author} {\bibfnamefont {J.~R.}\ \bibnamefont {Prance}}, \bibinfo
  {author} {\bibfnamefont {D.}~\bibnamefont {Kim}}, \bibinfo {author}
  {\bibfnamefont {J.~K.}\ \bibnamefont {Gamble}}, \bibinfo {author}
  {\bibfnamefont {R.~T.}\ \bibnamefont {Mohr}}, \bibinfo {author}
  {\bibfnamefont {Z.}~\bibnamefont {Shi}}, \bibinfo {author} {\bibfnamefont
  {D.~E.}\ \bibnamefont {Savage}}, \bibinfo {author} {\bibfnamefont {M.~G.}\
  \bibnamefont {Lagally}}, \bibinfo {author} {\bibfnamefont {M.}~\bibnamefont
  {Friesen}}, \bibinfo {author} {\bibfnamefont {S.~N.}\ \bibnamefont
  {Coppersmith}}, \ and\ \bibinfo {author} {\bibfnamefont {M.~A.}\ \bibnamefont
  {Eriksson}},\ }\href {\doibase 10.1073/pnas.1412230111} {\bibfield  {journal}
  {\bibinfo  {journal} {Proceedings of the National Academy of Sciences}\
  }\textbf {\bibinfo {volume} {111}},\ \bibinfo {pages} {11938} (\bibinfo
  {year} {2014})},\ \Eprint {http://arxiv.org/abs/1403.0019} {arXiv:1403.0019}
  \BibitemShut {NoStop}%
\bibitem [{\citenamefont {Foletti}\ \emph {et~al.}(2009)\citenamefont
  {Foletti}, \citenamefont {Bluhm}, \citenamefont {Mahalu}, \citenamefont
  {Umansky},\ and\ \citenamefont {Yacoby}}]{qubit4}%
  \BibitemOpen
  \bibfield  {author} {\bibinfo {author} {\bibfnamefont {S.}~\bibnamefont
  {Foletti}}, \bibinfo {author} {\bibfnamefont {H.}~\bibnamefont {Bluhm}},
  \bibinfo {author} {\bibfnamefont {D.}~\bibnamefont {Mahalu}}, \bibinfo
  {author} {\bibfnamefont {V.}~\bibnamefont {Umansky}}, \ and\ \bibinfo
  {author} {\bibfnamefont {A.}~\bibnamefont {Yacoby}},\ }\href {\doibase
  10.1038/nphys1424} {\bibfield  {journal} {\bibinfo  {journal} {Nature
  Physics}\ }\textbf {\bibinfo {volume} {5}},\ \bibinfo {pages} {903} (\bibinfo
  {year} {2009})}\BibitemShut {NoStop}%
\bibitem [{\citenamefont {Maze}\ \emph {et~al.}(2011)\citenamefont {Maze},
  \citenamefont {Gali}, \citenamefont {Togan}, \citenamefont {Chu},
  \citenamefont {Trifonov}, \citenamefont {Kaxiras},\ and\ \citenamefont
  {Lukin}}]{NV-group-theory}%
  \BibitemOpen
  \bibfield  {author} {\bibinfo {author} {\bibfnamefont {J.~R.}\ \bibnamefont
  {Maze}}, \bibinfo {author} {\bibfnamefont {A.}~\bibnamefont {Gali}}, \bibinfo
  {author} {\bibfnamefont {E.}~\bibnamefont {Togan}}, \bibinfo {author}
  {\bibfnamefont {Y.}~\bibnamefont {Chu}}, \bibinfo {author} {\bibfnamefont
  {A.}~\bibnamefont {Trifonov}}, \bibinfo {author} {\bibfnamefont
  {E.}~\bibnamefont {Kaxiras}}, \ and\ \bibinfo {author} {\bibfnamefont
  {M.~D.}\ \bibnamefont {Lukin}},\ }\href {\doibase
  10.1088/1367-2630/13/2/025025} {\bibfield  {journal} {\bibinfo  {journal}
  {New Journal of Physics}\ }\textbf {\bibinfo {volume} {13}},\ \bibinfo
  {pages} {025025} (\bibinfo {year} {2011})}\BibitemShut {NoStop}%
\bibitem [{\citenamefont {Rayson}\ and\ \citenamefont
  {Briddon}(2008)}]{zfs-derivation}%
  \BibitemOpen
  \bibfield  {author} {\bibinfo {author} {\bibfnamefont {M.~J.}\ \bibnamefont
  {Rayson}}\ and\ \bibinfo {author} {\bibfnamefont {P.~R.}\ \bibnamefont
  {Briddon}},\ }\href {\doibase 10.1103/PhysRevB.77.035119} {\bibfield
  {journal} {\bibinfo  {journal} {Phys. Rev. B}\ }\textbf {\bibinfo {volume}
  {77}},\ \bibinfo {pages} {035119} (\bibinfo {year} {2008})}\BibitemShut
  {NoStop}%
\end{thebibliography}%

\end{document}